\documentclass[12pt]{article}

\usepackage{a4wide}

\usepackage{latexsym}
\usepackage{epsf}
\usepackage{amssymb}
\usepackage{graphicx}
\usepackage{amsmath, cite}
\usepackage{amsmath,amssymb,amsthm}
\usepackage{verbatim}
\usepackage{hyperref}
\usepackage{amsmath}
\usepackage{amsfonts}
\usepackage{xfrac}
\usepackage{slashed}
\usepackage{cancel}
\usepackage{bbm}
\usepackage{bbold}

\usepackage{footmisc}

\usepackage[utf8]{inputenc}

\usepackage{fancyhdr}
\usepackage{datetime}

\usepackage{makeidx}

\usepackage{pgfplots}

\renewcommand{\Re}{\textrm{Re}}

\newcommand\varpm{\mathbin{\vcenter{\hbox{%
  \oalign{\hfil$\scriptstyle+$\hfil\cr
          \noalign{\kern-.3ex}
          $\scriptscriptstyle({-})$\cr}%
}}}}

\newcommand\varmp{\mathbin{\vcenter{\hbox{%
   \oalign{\hfil$\scriptstyle-$\hfil\cr
           \noalign{\kern-.3ex}
          $\scriptscriptstyle({+})$\cr}%
}}}}

  \def\cO{{\cal O}}

\def\beq{\begin{equation}}
\def\eeq{\end{equation}}

\begin{document}

\begin{titlepage}
\begin{center}
\rightline{\small }

\begin{flushright}
\end{flushright}

\vskip 2cm

{\Large \bf Algorithmically solving the Tadpole Problem }
\vskip 1.2cm

{ Iosif Bena$^{a}$, Johan Bl{\aa}b{\"a}ck$^{b}$, Mariana Gra\~na$^{a}$ and
Severin L\"ust$^{c,d}$ }
\vskip 0.1cm
{\small\it  $^{a}$ Institut de Physique Th\'eorique,
Universit\'e Paris Saclay, CEA, CNRS\\
Orme des Merisiers \\
91191 Gif-sur-Yvette Cedex, France} \\
\vskip 0.1cm
{\small\it  $^{b}$ Dipartimento di Fisica, Universit\`a di Roma ``Tor Vergata" \& INFN - Sezione di Roma2 \\
Via della Ricerca Scientifica 1, 00133 Roma, Italy} \\
\vskip 0.1cm
{\small\it  $^{c}$ Jefferson Physical Laboratory, Harvard University, Cambridge, MA 02138, USA }
\vskip 0.1cm
{\small\it  $^d$ Centre de Physique Th\'eorique, Ecole Polytechnique, CNRS \\
91128 Palaiseau Cedex, France} \\
\vskip 0.8cm

{\tt }

\end{center}

\vskip 1cm

\begin{center} {\bf Abstract }\\
\end{center}

The extensive computer-aided search applied in \cite{Bena:2020xrh} 
to find the minimal charge sourced by the fluxes that stabilize all the (flux-stabilizable) moduli of a smooth K3$\times$K3 compactification uses differential evolutionary algorithms supplemented by local searches.
We present these algorithms in detail and show that they can also solve our minimization problem for other lattices.
Our results support the Tadpole Conjecture: The minimal charge grows linearly with the dimension of the lattice and, for K3$\times$K3, this charge is larger than allowed by tadpole cancelation.

Even if we are faced with an NP-hard lattice-reduction problem at every step in the minimization process, we find that differential evolution is a good technique for identifying the regions of the landscape where the fluxes with the lowest tadpole can be found. We then design a ``Spider Algorithm," which is very efficient at exploring these regions and producing large numbers of minimal-tadpole configurations.

\vspace{0.2cm}

\end{titlepage}

\tableofcontents

\vspace{1.0cm}

\section{Introduction}\label{sec:intro}\label{sec:introduction}

Compactifications of String or M-theory  generically come with large numbers of (unphysical) massless scalar fields corresponding to the K\"ahler and complex-structure moduli of the compactification manifold. Many of these moduli can be given a mass by turning on topologically non-trivial magnetic fluxes along the cycles of the internal geometry. However, these fluxes also source electric brane charges, which must sum up to zero on a compact manifold. Hence, brane tadpole-cancellation conditions place upper bounds on the amount of these fluxes. Furthermore, since fluxes are integer quantized, there is a  lower bound on their (non-zero) quantum numbers.  It is therefore not clear whether there always exists a choice of fluxes within the tadpole bound that lead to a four-dimensional vacuum in which all the flux-stabilizable moduli have acquired a mass.

In a previous paper we have argued that whenever the number of moduli is large, one cannot stabilize all moduli within the tadpole bound \cite{Bena:2020xrh}. More precisely, we formulated the {\bf Tadpole Conjecture:} {\em The fluxes that stabilize a large number, $n$, of moduli at a generic point in moduli space, has a positive contribution to the tadpole-cancelation condition that grows at least linearly with the number of moduli:}
\beq
Q^{\rm stabilization}> \alpha \times n.
\eeq
Based on several explicit examples we also conjectured that
$
\alpha >  \frac13
$.
On the other hand, in M and F-theory compactifications on Calabi-Yau four-folds,
the negative contribution to the tadpole cancelation condition, that we call $-Q^{\rm loc}$, behaves as
$
Q^{\rm loc} \sim \frac14 n
$,
when the number of complex-structure moduli is large (compared to the number of K\"ahler moduli). This can be seen easily by noting that $Q^{\rm loc}=\frac{\chi}{24}\sim \tfrac14 h^{3,1}$ for large $h^{3,1}$.\footnote{In Type IIB language, most of this negative contribution  comes, in the large-tadpole limit, from D7  branes wrapping highly curved four-cycles on the base manifold.}
Our conjecture, if true, would rule out all compactifications with a large number of  stabilized moduli.

In \cite{Bena:2020xrh} we presented several examples supporting this conjecture: compactifications of F-theory on the sextic Calabi-Yau four-fold \cite{Braun:2020jrx}, Type IIB on Calabi-Yau manifolds stabilized at symmetric points in moduli space \cite{Giryavets:2003vd,Demirtas:2019sip}, Type IIB compactifications with D7-branes on ${\mathbb CP}^3$ \cite{Collinucci:2008pf},  and M-theory on K3$ \times $K3 \cite{Bena:2020xrh}. The latter has been the very first example of moduli stabilization \cite{Dasgupta:1999ss} and has been extensively studied, most notably in \cite{Aspinwall:2005ad} (see \cite{Bena:2020xrh} for more references). However, these works have concentrated on stabilization at specific points in moduli space where the compactification manifold supports matter with non-Abelian gauge symmetries.
While these non-Abelian gauge symmetries are desirable from the point of view of phenomenology, we believe that including the ``singularities" that give rise to this matter is a lose-lose game as far as tadpole-cancelation conditions are concerned:  they host a number of massless fields that is quadratic in the rank of the gauge group, and if one turns on fluxes to stabilize all of them we expect  the tadpole to be also quadratic in the rank of the gauge group. We therefore required in \cite{Bena:2020xrh} that all moduli be stabilized away from points where there are are non-Abelian gauge groups.

The K3$\times$K3 compactification is particularly amenable to computer-based searches of the fluxes that stabilize all the moduli. This is because of the results of \cite{Braun:2008pz}, where  the condition of full moduli stabilization is translated into a set of requirements  on the eigenvalues of an integer matrix involving the square of the fluxes. Explicitly, the M-theory four-form flux has two legs on each K3, and thus the flux numbers are encoded in a $22 \times 22$ matrix, $N$.\footnote{Remember that K3 has 22 homologically-different two-cycles.} The complete moduli stabilization is achieved when the matrix $N d N^t d$ (where $d$ is the intersection matrix of K3) has non-negative eigenvalues, up to some subtleties that we present in detail in Section \ref{sec:tadpole_problem}. Furthermore, the charge induced by the fluxes is just the trace of this matrix.
It was already demonstrated in \cite{Dasgupta:1999ss, Aspinwall:2005ad,Braun:2008pz} that there exist valid flux configurations within the tadpole bound, but all these examples have either non-Abelian gauge groups  or do not have all moduli stabilized by fluxes. Therefore, our systematic search aims at understanding if this behavior is generic, or just an artifact of a special choice of easily constructible examples.

The goal of our search is to find flux matrices with the lowest possible tadpole and which give rise to a smooth compactification where all the flux-stabilizable moduli are stabilized. To find such matrices we utilize two search algorithms: a global, population-based search called {\it differential evolution}, and a local search we have chosen to call {\it the Spider}.
Differential evolution (DE) \cite{Storn:1996,Storn:1997,Price:2005,Engelbrecht} can be used for problems where we can establish some distance to our desired target in terms of a ``cost'' or ``fitness'' function. The DE algorithms  make combinations of different individuals in a population (here the individuals are the flux matrices) to form new ones, progressively improving their fitness. Once the DE appears to converge on a minimum of the fitness function, we explore this population further with the Spider. The Spider takes the members of the population, and slightly changes them generating matrices with one to two entries changed per iteration. As shown in detail in Table 2 in the main text, with this further local search we are able to get many more matrices with minimal tadpole and, furthermore, in most of the examples this algorithm finds matrices with lower tadpole than those found by the DE algorithm.

Similar evolutionary or genetic algorithms were previously used to explore the string landscape and flux compactifications in \cite{Blaback:2013ht, Damian:2013dq, Damian:2013dwa, Blaback:2013fca, Blaback:2013qza, Abel:2014xta, Ruehle:2017mzq, Cole:2019enn,AbdusSalam:2020ywo,CaboBizet:2020cse}.
Another computer-aided search of flux vacua constrained by tadpole cancellation was performed in \cite{Betzler:2019kon}.
For other uses of data science and machine-learning techniques in the context of String Theory see \cite{He:2017aed,Carifio:2017bov, Carifio:2017nyb,Hashimoto:2018ftp,Wang:2018rkk,Bull:2018uow,Demirtas:2018akl,Constantin:2018hvl,Klaewer:2018sfl,Halverson:2018cio,Mutter:2018sra,Altman:2018zlc,He:2018jtw,Cole:2018emh,Bull:2019cij,Halverson:2019kna,Hashimoto:2019bih,Halverson:2019tkf,He:2019vsj,Brodie:2019dfx,Ashmore:2019wzb,Parr:2019bta,Halverson:2019vmd,Halverson:2020opj,Ruehle:2020jrk,Parr:2020oar,Otsuka:2020nsk,Deen:2020dlf,Krippendorf:2020gny,He:2020eva,Akutagawa:2020yeo,Bao:2020nbi,He:2020bfv,Bies:2020gvf,Erbin:2020srm,Erbin:2020tks,Demirtas:2020dbm,He:2020lbz,Parr:2020dbl,Cole:2020ktw,Jejjala:2020wcc}.

While it may seem natural to approach this question by computer-based searches, the problem we are solving is still very hard. With a computer we can explore the space of flux matrices, however, this approach is complicated by our requirement that all moduli be stabilized away from singularities.\footnote{This condition is satisfied if there are no integer eigenvectors of norm minus two orthogonal to the eigenvectors of positive norm (see details in Section \ref{sec:tadpole_problem}).}
Determining if a flux matrix, $N$, satisfies this requirement is equivalent to finding the shortest non-zero vector in a certain integer lattice.
This so-called ``shortest-vector problem'' (SVP) is known to be NP-hard \cite{proceedingsAjtai:1997,proceedingsAjtai:1998}. This means we have to solve an NP-hard problem just for verifying each candidate that is generated during the minimization problem. Hence, at least in the approach we have taken here, the problem appears to be harder than an NP-problem.\footnote{An NP-class problem is one in which a candidate solution can be verified in polynomial time, thus our problem appears to be beyond this class.}

To get around this issue, we have taken advantage of the Lenstra-Lenstra-Lov\'asz (LLL) algorithm \cite{Lenstra82factoringpolynomials}, which gives a good approximate solution for the shortest-vector problem and is guaranteed to complete it in polynomial time. Its only draw-back is that it may give false-positives,\footnote{A false-positive means that the algorithm failed to find the shortest vector, and when that happens we cannot exclude the corresponding flux matrix during the search (only post-search). The rate at which we can verify matrices post-search and eliminate false-positives depends on the hardware; on our table-top machines we could verify about 1000 matrices per second.} which have to be excluded by hand. We have not analyzed the complexity of the implementation of our modified LLL algorithm, but are satisfied with the speed at witch we verify matrices and the moderate amount of false-positives produced.

In this paper we present the gory details of this search procedure. In order to check the robustness of our conclusions, we generalized the problem from the K3-lattice $H^2(K3, \mathbb{Z})$ to arbitrary even lattices, $\Lambda$, with mixed signature. This allows us to analyze a similar minimization problem using smaller $D \times D$ matrices, with $6 \leq D = \dim(\Lambda) \leq 22$. We present these results in full detail.

From our searches, summarized in Table~\ref{tab:summary}, we conclude that if the conditions analogous to moduli stabilization and smoothness are satisfied, there is always a minimal flux-induced charge, depending on the lattice in question, which scales like
\beq
Q_{\rm min}\sim D,
\eeq
where $D$ is the dimension of the lattice.
In particular, for K3$\times$K3, we find the minimal charge to be 25, which is larger than 24, the maximal valued allowed by tadpole cancelation.

The final version of the code that we have used in this paper and in \cite{Bena:2020xrh} has been made publicly available online at as part of the \texttt{bbsearch} project \cite{bbsearch} (we refer to the sections called ``minitad'' in that project). It includes the fitness function we developed, the Spider algorithm, various utility tools, and an extensive documentation on how to use \texttt{bbsearch}.
\vspace{0.1cm}

This paper is organized as follows. In Section 
\ref{sec:K3K3} we summarize the problem we want to solve and which was briefly presented in {\cite{Bena:2020xrh}}. We continue in Section \ref{sec:problem} to show how the problem can be extrapolated to smaller matrices, where our search works even more reliably and offers further support to our result for K3$\times$K3. In Section \ref{sec:algorithms} we explain the approach we have used: Designing a fast fitness function to sort out appropriate matrices, which is then used for a differential evolution search and later for the {\it Spider} local search. In Section~\ref{sec:toy} we present the results of these searches. In Section \ref{sec:discussion} we discuss these results and present some concluding remarks.
There are two appendices where we present certain mathematical details underlying our algorithms.

\begin{table}[hbt]
\centering
\begin{tabular}{cc}

\begin{tabular}{c|c|c}\label{tab:summary}
lattice $\Lambda$ & $D\!=\!\dim(\Lambda) $ & $Q_\mathrm{min}(\Lambda)$\\
\hline
$3\,  U $ & 6  & 5 \\
    $A_4 \oplus U$ & 6  & 6 \\
    $D_4 \oplus U$ & 6 &  6   \\
$A_4 \oplus 2 \, U $  & 8 &  7 \\
$D_4 \oplus 2 \, U$  & 8 &  6 \\
$E_6 \oplus U$ & 8  & 9  \\
$A_4 \oplus 3 \, U$  & 10 &  9  \\
$D_4 \oplus 3\,  U$  & 10 & 9  \\
\end{tabular}

\hspace{1em} & \hspace{1em}

\begin{tabular}{c|c|c}
lattice $\Lambda$ & $D\!=\!\dim(\Lambda) $ & $Q_\mathrm{min}(\Lambda)$ \\
\hline
    $E_8 \oplus U$ & 10    & 10 \\
    $E_8 \oplus 2 \,  U$ & 12 &  12 \\
    $E_8 \oplus 3\,  U$ & 14 &  13 \\
    $2 \, E_6 \oplus 2\,  U $  &16 &  14  \\
     $2 \, E_8 \oplus U $ & 18 &   20 \\
     $ 2\, E_8 \oplus 2\,  U $ & 20 &  21  \\
     $2 \, E_8 \oplus 3\,  U$ & 22 &  25 \\
     && \\
\end{tabular}
\end{tabular}
\caption[Overview of the results.]{\it An overview of all root lattices%
\footnotemark \,
that we processed using the algorithms described in Section~\ref{sec:algorithms}.
We list the lattices, their dimensions and the minimal values of $Q$ found.
A detailed account of our results can be found in Section~\ref{sec:toy}.}
\end{table}

\footnotetext{Recall that $A_n$, $D_n$ and $E_n$ denote the root lattices of the Lie algebras of the special unitary groups $SU(n+1)$, the special orthogonal groups $SO(2n)$ and the exceptional groups, respectively.
Moreover, $U$ is the two-dimensional lattice with quadratic form of signature (1,1), given below Equation \eqref{intK3}.}

\newpage

\section{The Tadpole Problem}\label{sec:tadpole}\label{sec:tadpole_problem}

Compactifications of M-theory or F-theory on Calabi-Yau four-folds $CY_4$,
have $h^{1,1}$ K\"ahler moduli and $h^{3,1}$ complex-structure moduli.
The latter can be stabilized by turning on topologically-nontrivial components of the four-form field strength, $G_4$, along the internal directions.
These fluxes must satisfy the tadpole-cancellation condition
\begin{equation}\label{eq:generaltadpole}
\frac12 \int_{CY_4} G_4 \wedge G_4 + N_{\rm M2/D3} = \frac{\chi(CY_4)}{24} \,,
\end{equation}
where $N_{\rm M2/D3}$ denotes the number of space-time filling M2/D3 branes and $\chi(CY_4)$ is the Euler number of the four-fold, which can be expressed in terms of the Hodge numbers:
\begin{equation}
\chi(CY_4) = 6(8 + h^{1,1} + h^{3,1} - h^{2,1}) \,.
\end{equation}
Hence, the right-hand side of \eqref{eq:generaltadpole} scales linearly with the number of complex-structure moduli, $h^{3,1}$. In the limit where $h^{3,1} \gg h^{1,1}$ and $h^{3,1} \gg h^{2,1}$ one has
\begin{equation}
\frac{\chi(CY_4)}{24} \sim \frac{h^{3,1}}{4} \,.
\end{equation}
We conjectured that for every integer four-form flux that stabilizes the moduli of a smooth compactification, the integral on the left-hand side of \eqref{eq:generaltadpole} satisfies the bound
\begin{equation}
\frac12 \int_{CY_4} G_4 \wedge G_4 > \alpha h^{3,1} \,,
\end{equation}
where $\alpha$ is an $\cO(1)$-constant.
Notice, that when $\alpha > 1/4$ it is impossible to construct compactifications with large $h^{3,1}$ and all moduli stabilized.
In \cite{Bena:2020xrh} we have also formulated the stronger conjecture that $\alpha > \frac13$.

\subsection{K3 $\times$ K3}
\label{sec:K3K3}

For M-theory compactified on the product manifold $CY_4 = \mathrm{K3} \times \mathrm{K3}$, moduli stabilization is well understood, and can be formulated exclusively in terms of integer matrices, their eigenvalues and their eigenvectors \cite{Braun:2008pz, Braun:2010ff}. In particular, there is no need to have any knowledge of complicated period integrals of the Calabi-Yau manifold, or the corresponding Picard-Fuchs equations. This makes these compactifications well suited for algorithmic searches.

We start by reviewing the relevant results of \cite{Braun:2008pz, Braun:2010ff}:
Four-form fluxes $G_4 \in H^4(K3 \times K3, \mathbb{Z})$ can have components whose four legs lie entirely in one of the two K3's, as well as components that have two legs on one K3 and two legs on the other. Here, we only consider the second option:
\begin{equation}\label{eq:K3K3flux}
G_4 \in H^2(K3, \mathbb{Z}) \times H^2(\widetilde{K3}, \mathbb{Z}) \,.
\end{equation}
For these fluxes the vacuum condition,
\begin{equation}\label{eq:selfdual}
G_4 = \ast G_4 \,,
\end{equation}
can stabilize all geometric moduli (K\"ahler and complex structure) of the two K3's, except for their volumes.

The middle cohomology $H^2(K3, \mathbb{Z})$ has the natural inner product
\begin{equation}\label{eq:K3innerprod}
(\alpha, \beta) = \int_{K3} \alpha \wedge \beta \,.
\end{equation}
In terms of a suitable basis, $\alpha_i \in H^2(K3, \mathbb{Z})$ ($i = 1, \dots, 22$), the corresponding intersection matrix
\begin{equation}
d_{ij} = \int_{K3} \alpha_i \wedge \alpha_j
\end{equation}
is given by
\begin{equation} \label{intK3}
d = U \oplus U \oplus U \oplus (-E_8) \oplus (-E_8) \,.
\end{equation}
Here $E_8$ is the Cartan matrix of $E_8$, the matrix $U$ is defined as $U \equiv \begin{pmatrix}0 & 1 \\ 1 & 0 \end{pmatrix}$ and
$H^2(K3, \mathbb{Z})$ is the unique even self-dual lattice of signature $(3,19)$.
Moreover, a point in the moduli space of K3 corresponds to a choice of three self-dual 2-forms
\begin{equation}
\omega_a \in H^2(K3, \mathbb{R}) \,, \qquad a = 1,2,3 \,,
\end{equation}
which define a Hyper-K\"ahler structure on K3.
The self-duality of $\omega_a$ is equivalent to the positivity of their norm with respect to \eqref{eq:K3innerprod}.
Therefore $\omega_a$ span a 3-plane $\Sigma$ in $H^2(K3, \mathbb{R})$ such that the restriction of $d_{ij}$ to $\Sigma$ is positive definite.

The four-form flux \eqref{eq:K3K3flux} allows to define two homomorphisms $g \colon H^2(\widetilde{K3}) \rightarrow H^2(K3)$ and $\tilde g \colon H^2(K3) \rightarrow H^2(\widetilde{K3})$:
\begin{equation}
g(\tilde v) \equiv \int_{\widetilde{K3}} G_4 \wedge \tilde v \,,\qquad \tilde g(v) \equiv \int_{K3} G_4 \wedge v \,.
\end{equation}
Note that  $g$ and $\tilde g$ are adjoint with respect to the inner product \eqref{eq:K3innerprod}.
If we expand $G_4$ in terms of the integer bases
\begin{equation}
G_4 = N^{ij} \alpha_i \wedge \tilde \alpha_j \,,
\end{equation}
the two maps have the matrix representations
\begin{equation} 
g^i{}_j = N^{ik} d_{kj} \,, \qquad \tilde g^i{}_j = \left(N^T\right)^{ik} d_{kj} \,.
\end{equation}
It was shown in \cite{Braun:2008pz} that the vacuum condition \eqref{eq:selfdual} is equivalent to the following condition on $g$ and $\tilde g$:

\smallskip
\noindent \emph{The matrices $(g\tilde g)^i{}_j = N^{i \tilde k} d_{\tilde k \tilde l} N^{m\tilde l} d_{m j}$ and $(\tilde g g)^{\tilde \imath}{}_{\tilde \jmath} = N^{k \tilde \imath} d_{kl} N^{l\tilde m} d_{\tilde m \tilde \jmath}$ are diagonalizable with non-negative eigenvalues.}

When this condition is satisfied, the eigenvectors of $g\tilde g$ with positive norm span a three-plane, $\Sigma$, in $H^2(K3, \mathbb{R})$.
Therefore, they determine a point in the moduli space of K3.
Consequently, the question of moduli stabilization translates into the uniqueness of such a $\Sigma$.
To make this more specific, denote the eigenvalues of positive-norm and negative-norm eigenvectors by $\{a_1,a_2,a_3\}$ and $\{b_1, \dots, b_{19}\}$, respectively.
If there is no $a_i$ and $b_j$ such that $a_i = b_j$ all moduli are stabilized.
Otherwise the three-plane $\Sigma$ is not uniquely defined and some moduli remain massless.
We furthermore look for stabilization away from orbifold singularities. These arise when there is a root (an integer vector of norm-square -2, later also sometimes called a ``norm-minus-two'' vector) orthogonal to $\Sigma$.
Equivalent statements hold for the eigenvectors of  $\tilde g g$ and the moduli space of $\widetilde{\mathrm{K3}}$.

Finally, it is not difficult to see that the M2-brane charge sourced by the fluxes, which enters in the tadpole-cancellation condition \eqref{eq:generaltadpole}, is;
\begin{equation}
\frac12 \int G_4 \wedge G_4 = \tfrac12 N^{i \tilde k}  d_{\tilde k \tilde l} N^{n\tilde l} d_{n i} = \tfrac12 \mathrm{tr} \, g \tilde g \,.
\end{equation}

\subsection{Generalization to arbitrary lattices}\label{sec:problem}

Inspired by the matrix formulation of moduli stabilization on K3$ \times $K3 we can investigate a slightly more general problem.
Consider an even lattice, $\Lambda$, of dimension $D$ with basis $\{\alpha_i\}$ and inner product $(\cdot, \cdot)$ of signature $(p, D-p)$.
With respect to this basis the inner product has the matrix representation
\begin{equation}\label{eq:latticeinner}
d_{ij} = (\alpha_i, \alpha_j) \,.
\end{equation}
Given moreover an integer matrix $\left(N^{ij}\right) \in \mathbb{Z}^{D\times D}$ (or equivalently $N \in \Lambda \otimes \Lambda$) we define a map $g\colon \Lambda \rightarrow \Lambda$ and its adjoint $\tilde g$ by
\begin{equation}\label{eq:gmatrix}
g^i{}_j = N^{ik} d_{kj} \,, \qquad \tilde g^i{}_j = N^{ki} d_{kj} \,.
\end{equation}
Let us assume that $N$ satisfies the following conditions
\begin{itemize}
\item The maps $g \tilde g$ and $\tilde g g$ are diagonalizable over $\mathbb{R}$ with non-negative eigenvalues.%
\footnote{When referring to eigenvalues / eigenvectors of $g \tilde g$ and $\tilde g g$, we always implicitly consider eigenvalues and eigenvectors of the induced maps $g \tilde g, \tilde g g \colon \Lambda \otimes \mathbb{R} \rightarrow \Lambda \otimes \mathbb{R}$.
Generically there will be no integer eigenvalues or eigenvectors in $\Lambda$.}
\item If there is a degenerate eigenvalue, the restriction of $d_{ij}$ to the corresponding eigenspace is of definite signature.
\item There is no root $\alpha \in \Lambda$ (or equivalently no lattice element with $\| \alpha \|^2 = \pm 2$) such that $\alpha$ is orthogonal to all positive-norm eigenvectors of $g \tilde g$ or to all positive-norm eigenvectors of $\tilde g g$.
\end{itemize}
We introduce the quantity
\begin{equation}\label{eq:Q}
Q_\mathrm{min}(\Lambda) \equiv \min Q(N) \,,\qquad\text{with}\qquad Q(N) \equiv \frac12 \mathrm{tr} (g \tilde g) \,,
\end{equation}
where the minimum is taken over all matrices $N$ such that the above three conditions are satisfied.
It is immediately clear that $Q_\mathrm{min}(\Lambda) > 0$, because if there is an $N$ with $Q(N) \leq 0$ the first condition implies $g \tilde g = \tilde g g = 0$ and therefore the second condition would not be satisfied.

In the remainder of the paper we will use differential evolutionary algorithms to determine $Q_\mathrm{min}(\Lambda)$ for a set of example lattices, $\Lambda$.
Even though this approach can strictly speaking only give an upper bound on $Q_\mathrm{min}(\Lambda)$, all our results indicate that
\begin{equation}
Q_\mathrm{min}(\Lambda) \sim D \, ,
\end{equation}
where $D$ is the dimension of $\Lambda$ (a summary of the lattices that we analyzed can be found in Table~\ref{tab:summary}). It is therefore tempting to conjecture that this relation holds universally.

Before going into the details of the algorithmic approach, a few more comments are in order.
As one can see from the definitions \eqref{eq:gmatrix}, $\tilde g$ is the adjoint of $g$ with respect to the inner product \eqref{eq:latticeinner}.
Therefore, both $g \tilde g$ and $\tilde g g$ are self-adjoint.
On spaces with a positive-definite inner product, self-adjoint matrices of the form $A^\dagger A$ are always diagonalizable with non-negative eigenvalues.
This does not happen for indefinite bilinear forms and therefore the first condition in the list above is non-trivial.

There are $p$ eigenvectors of $g \tilde g$ of positive norm-square.
Let us denote their span by $\Sigma$ (and $\tilde \Sigma$ for the eigenvectors of  $\tilde g g$).
The second condition ensures that $\Sigma$ and $\tilde \Sigma$ are uniquely defined $p$-dimensional subspaces of $\Lambda \otimes \mathbb{R}$.
Therefore the third condition is always well-defined.
Moreover, the decomposition of $\Lambda \otimes \mathbb{R}$ into $\Sigma$ and its orthogonal complement $\Sigma^\perp$ brings $g \tilde g$ into block-diagonal form.
Both blocks act on spaces whose inner product has a definite signature and therefore share all the familiar properties of normal self-adjoint maps.
In particular, $g$ and $\tilde g$ are also block diagonal and satisfy $g(\tilde \Sigma) \subseteq \Sigma$ and
$\tilde g(\Sigma) \subseteq \tilde\Sigma$.
This last property was used in \cite{Braun:2008pz} to show that any M-theory compactification on K3$\times$K3 with a flux matrix satisfying the first and second condition has a Minkowski vacuum with stabilized moduli.

The existence of a root orthogonal to $\Sigma$ requires that $\Lambda \cap \Sigma^\perp \neq 0$, where $\Sigma^\perp$ is spanned by the $D-p$ eigenvectors with negative norm-square.
The components of the eigenvectors are solutions of degree-$D$ algebraic equations, and are generically not integer.
Therefore, a linear combination of only $D-p$ or less eigenvectors will generically not be in $\Lambda$. This can only happens if the characteristic polynomial of $g \tilde g$ is reducible over the integers.
Following the more detailed discussion in Appendix~\ref{app:integervectors}, the conditions on the existence of a root in $\Sigma^\perp$ can be summarized as follows:
\begin{enumerate}
\item Factorize the characteristic polynomial $p(\lambda) = \det(g \tilde g - \lambda \mathbb{1})$ over the integers,
\begin{equation}
p(\lambda) = \prod_r p_r(\lambda)^{m_r} \,.
\end{equation}
\item For each irreducible factor $p_r(\lambda)$ compute
\begin{equation}
N_r \equiv \ker p_r(g \tilde g) \,.
\end{equation}
Notice that this denotes the kernel in $\Lambda$.
Since $g \tilde g$ is diagonalizable, $\dim(N_r) = m_r\, \mathrm{rank} \, p_r (g \tilde g)$ and $\Lambda = \bigoplus_r N_r$ is an orthogonal decomposition of $\Lambda$.
\item If the restriction of the inner product to $N_r$ is negative definite, $N_r$ is orthogonal to $\Sigma$. On the other hand, if the inner product on $N_r$ is positive definite or indefinite, it contains at least one eigenvector of positive norm-square; however any integer vector in $N_r$ is a linear combination of all eigenvectors in $N_r \otimes \Lambda$, so it cannot be orthogonal to this negative-norm eigenvector.\footnote{There is a further subtlety if the multiplicity of $p_r$ is larger than one. It is not hard to see that when this happens the second condition guarantees also that $N_r$ can only contain an integer vector orthogonal to $\sigma$ if it is positive definite.}
Therefore,
\begin{equation}\label{eq:Nminus}
N_{-} \equiv \bigoplus \left\{ N_r :  \text{$\left.(\cdot, \cdot)\right|_{N_r}$ is negative definite} \right\} = \Lambda \cap \Sigma^\perp \,.
\end{equation}
\end{enumerate}
Consequently, any root orthogonal to $\Sigma$ must be in $N_-$.
Since $N_-$ is a sublattice of $\Lambda$, the question of the existence of such roots boils down to finding the shortest possible lattice vector in $N_-$.
This is a well known problem and known to be NP-hard \cite{proceedingsAjtai:1997,proceedingsAjtai:1998}.

\section{The Algorithms}\label{sec:algorithms}

In this section we describe our algorithmic approach to the problem introduced in the previous section.
We start by outlining the basic principles of Differential Evolution in Section~\ref{sec:DE}, before we discuss in detail the design of a suitable fitness function in Section~\ref{ssec:fitness}.
In Section~\ref{sec:spider} we introduce a complementary local search algorithm which will be used to further optimize the results of the Differential Evolutionary search.

\subsection{Differential evolution}\label{sec:DE}

Differential Evolution (DE) is a collection of algorithms that are part of the subject of Evolutionary Computation, as the algorithms are inspired by the biological processes of evolution \cite{Storn:1996,Storn:1997,Price:2005,Engelbrecht}. The common feature of these algorithms is to form a population of candidate solutions and have the candidates interact to produce better candidates, to solve global optimization problems.

A DE algorithm works in the following way. Parametrize the problem such that a candidate solution can be represented by a vector $x \in \mathbb{R}^d$, and form a population of $M$ random candidates $\{x_a\}_{a=1,\ldots, M}$. Design a \emph{fitness function}, $\texttt{fitness} \colon \mathbb{R}^d \rightarrow \mathbb{R}$, which quantifies how well a candidate solution solves the problem.
The objective is to find the global minima of the fitness function.
 For example, when solving a system of algebraic equations the fitness function can be taken to be the sum of the absolute value squared of each equation. Select a particular member of this population, $x$, together with $N$ pairs of members distinct from $x$; $\{(y_\alpha,z_\alpha)\}_{\alpha=1,\ldots, N}$. A \emph{mutation operation} is then performed on $x$ to form a new vector $x'$ according to
\begin{equation}\label{eq:DEmut}
  x' = x + \sum_{\alpha=1}^{N} F^\alpha (y_\alpha - z_\alpha)\,,
\end{equation}
where the $F^\alpha$ are positive numbers. The new vector $x'$ and the old vector $x$ is then selected for a \emph{crossover operation}, e.g.
\begin{equation}
  {x''}_i = \left\{ \begin{array}{ll} {x'}_i & {\texttt{if rand()}} < C_r \\ x_i \end{array} \right. \,,
\end{equation}
where $C_r$ is some positive number less than $1$ called the \emph{crossover rate}. Finally $x$ and $x''$ are compared to each other, by evaluating how well they solve the problem at hand; that is by evaluating the fitness function for both vectors (or other vectors in the population), and we select
\begin{equation}
  x'' \texttt{ if fitness($x''$) < fitness($x$) else } x\,.
\end{equation}
The algorithm outlined above would be denoted as DE/rand/$N$/bin, as the entries selected for crossover are taken randomly, we have selected $N$ pairs to produce the mutation, and a binomial operation was used in the crossover operation.

When it comes to selecting a new population for the next generation, there are also a set of selection operations that are available. The implementation we use employs a \emph{tournament} selection, which means that out of all candidates a subset is repeatedly taken from which candidates are selected based on best fitness within the subset.

For our problem we did not implement our own DE algorithm, but instead we used the \emph{adaptive} DE/rand/1/bin \emph{with radius limited sampling} method of \texttt{BlackBoxOptim.jl} \cite{Feldt2018}. The adaptive feature means that search parameters are changed dynamically throughout the search, and a limited radius sampling is where only candidates that are sufficiently adjacent are compared.

We would also like to mention that we have used, in addition to \texttt{BlackBoxOptim.jl}, \texttt{bbsearch.jl} \cite{bbsearch} which is a form of user interface for \texttt{BlackBoxOptim.jl}. The fitness function we have designed to solve our problem, as outlined in Subsection \ref{ssec:fitness}, is hosted online as a part of the \texttt{bbsearch.jl} project.

Before moving on to the next section we would first like to introduce some nomenclature: The fitness will be written as a \emph{weighted sum} of what we will call \emph{penalties}. When all \emph{relevant} penalties are zero, we have a good candidate matrix. We will go through our penalties in the next section, but what we here call a \emph{relevant} penalty is one that is \emph{not} the penalty associated to the tadpole charge \eqref{eq:Q}, as we do not a priori know its minimum.

\subsection{The fitness function for the Tadpole Problem}\label{ssec:fitness}

We have chosen to attack the problem introduced in Section~\ref{sec:problem} using a differential-evolution algorithm.
As explained above, such an algorithm seeks to find the global minima of a fitness function. We therefore have to reformulate this problem into the design of a fitness function whose global minima are our desired solutions.

The problem consists of finding integer-valued matrices $N \in \mathbb{Z}^{D\times D}$ which satisfy a set of conditions and which minimize the quantity $Q(N)$ defined in \eqref{eq:Q}.
However, differential evolutionary algorithms operate much better on continuous data, therefore we work with a population consisting of $M$ initially random real $D \times D$ matrices.
Consequently, the fitness function will be a function $\texttt{fitness} \colon \mathbb{R}^{D\times D} \mapsto \mathbb{R}^+$.
To enforce the fact that the matrices are integer-valued, the first step in building the fitness function is to discretize the input, which we do by rounding the input data to the nearest integers.\footnote{There are two design paths one can take:\\ 1.~The DE can act on real-valued matrix entries in $\mathbb{R}^{D\times D}$, and the fitness function $\texttt{fitness}: \mathbb{R}^{D\times D} \mapsto \mathbb{R}^+$ can be used to enforce the fact that the input consists of real numbers\\ 2.~The DE can act on integer-valued matrix entries in $\mathbb{Z}^{D\times D}$ with fitness function $\texttt{fitness}: \mathbb{Z}^{D\times D} \mapsto \mathbb{R}^+$. \\The first option, which is the path we chose, contains more freedom for the algorithm to move. Consider the mutation (see Eq.~\ref{eq:DEmut}) of one entry when the DE acts on $\mathbb{R}^{D\times D}$: $0.4 \approx 0 \to 0.4 + 1*(0.9-0.7) = 0.6 \approx 1$, This mutation managed to switch the effective entry from $0$ to $1$, while the same DE acting on $\mathbb{Z}^{D\times D}$ with these numbers rounded, $0 \to 0 + 1*(1-1) = 0$, provides no evolution. 
}

The only free parameters in the specification of the problem are the entries of a matrix $d \in \mathbb{Z}^{D\times D}$, which defines the inner product in \eqref{eq:latticeinner} on an even lattice of signature $(p, D-p)$.
With the help of $d$ we define for each $N$ two more integer matrices $g \equiv N d$ and $\tilde g \equiv N^T d$, as in \eqref{eq:gmatrix}.
We have several goals that we want our fitness function to achieve:
\begin{enumerate}
  \item All eigenvalues of $g \tilde g$ are non-negative. \label{it:eig}
  \item $g \tilde g$ and $\tilde g g$ are both diagonalizable. \label{it:diag}
  \item If there are non-distinct eigenvalues, their eigenvectors must have norms of the same sign. \label{it:diff}
  \item   There is no vector with norm-squared equal to -2 (``norm-minus-two vector'') which is orthogonal to all $p$ eigenvectors (of  $g \tilde g$ or $\tilde g g$) of positive norm-square.\label{it:zero}
  \item Minimize the induced tadpole charge $Q(N) = \frac12\mathrm{tr}(g \tilde g)$. \label{it:tad}
\end{enumerate}
For each of the goals, we formulate one or more \emph{penalties}. A penalty is a function returning a non-negative number measuring how close the input matrix is to achieving that particular goal, zero meaning the goal has been reached. The final 
value that the fitness function will return is then computed as the weighted sum of each penalty
\begin{equation}
  \texttt{fitness}(N) = \sum_{k} w^k p_k(N)\,.
\end{equation}
$w^k$ representing the weights, and $p_k(N)$ the penalties, summed over all penalties we design.

To improve execution time, we only compute penalties as needed. The eigenvalue penalties of Item \ref{it:eig} and the norm-minus-two check, Item \ref{it:zero}, are always computed, but Item~\ref{it:diag} is only relevant if there are no non-distinct eigenvalues. Moreover, Item~\ref{it:diff} is only computed for diagonalizable matrices.

Before we continue we want to highlight the fact that the problem we are considering is very hard when it comes to computational complexity. To simply verify whether a matrix satisfies all the above goals we have to solve a lattice-reduction(--like) problem (Item \ref{it:zero} above), which is an NP-hard problem \cite{proceedingsAjtai:1997,proceedingsAjtai:1998}. Hence we cannot even verify if a candidate matrix satisfies our goals in polynomial time, with complete accuracy, and the problem we try to solve may not even be in the NP class of problems. It is therefore  important to choose algorithms that can solve this problem fast, even at the cost of generating false-positive candidates (meaning candidates that have a matrix that fails the condition in Item \ref{it:zero}, but that has zero penalty in our implementation). We will return to this point later. The final result of a search can then be analyzed by more accurate and slow methods, and false-positives can be filtered out.\footnote{This filtering is a Diophantine problem, which oftentimes (though not always) be solved by Mathematica. This can only be used for post-search verification, as the algorithms Mathematica uses too slow for running our search.}

We now proceed to describe how the we have designed the penalties. The implementation of the fitness function can be found as part of the \texttt{bbsearch.jl} project \cite{bbsearch}, which is an implementation in the Julia programming language \cite{Julia-2017}.\\

\begin{figure}
  \begin{center}
    \includegraphics{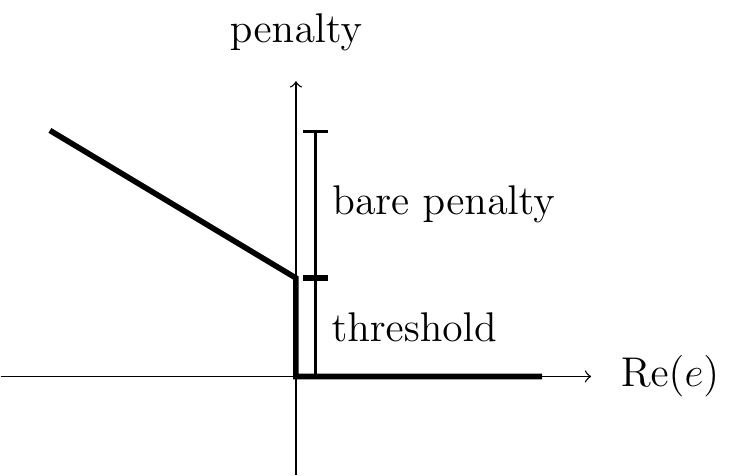}
  \end{center}
  \caption{\it An example of an effective penalty applied to eigenvalues: The bare penalty is the size of the negative real part of an eigenvalue, and a threshold value is added for the smallest non-zero penalty to avoid small unwanted numbers.}
  \label{fig:penalties}
\end{figure}

{\bf Item \ref{it:eig}: The goal of positive eigenvalues.} This goal is split into three separate penalties:
\begin{enumerate}
  \item The sum of the absolute value of the imaginary parts of all complex eigenvalues. \label{it:cplx_pen}
  \item The sum of the absolute value of negative real parts of all complex eigenvalues. \label{it:nc_pen}
  \item The sum of the absolute value of all negative real eigenvalues. \label{it:neg_pen}
\end{enumerate}
This means that when all eigenvalues are non-negative all these penalties are zero and the goal has been met. Each of these penalties are also supplemented with a \emph{threshold}, that is a constant value is added to the penalty when it is non-zero (accounting also for numerical zeroes). This is illustrated in Figure \ref{fig:penalties}. This is to prevent the algorithm to prefer very small negative eigenvalues together with very low tadpoles sourced by non-integer fluxes (that we will describe the penalty for later) over zero or positive eigenvalues with higher tadpoles.\\

{\bf Item \ref{it:diag}: The goal of diagonalizability.} There are a number of ways in which we can verify the diagonalizability of the flux matrix numerically, however what we found is that they are all too unstable to be used reliably. Our only option here is to compare the geometric and algebraic multiplicity.\footnote{The algebraic multiplicity denotes the multiplicity of an eigenvalues as the root of the characteristic polynomial.
The geometric multiplicity is the dimension of the corresponding eigenspace.
If the geometric multiplicity is smaller than the algebraic multiplicity for at least one eigenvalue, a matrix is not diagonalizable.}

For this we use \texttt{Nemo.jl} \cite{nemo} to compute and factorize the characteristic polynomial (which will also be used in Item \ref{it:zero}) to derive the geometric multiplicity. A constant penalty is then given if for any eigenvalue where the geometric and algebraic multiplicity do not match.
This method uses exact integer arithmetics and therefore does not suffer from numerical accuracy issues.
It is however slower than a purely numerical approach.
\\

{\bf Item \ref{it:diff}: The goal of distinct eigenvalues.} If some eigenvalues have algebraic multiplicity larger than one, the norms of their corresponding eigenvectors must have the same sign. We find that the most reliable way to check this numerically is by building a matrix for each eigenvalue
\begin{equation}
  b_{ij} = (v_i,v_j)
\end{equation}
using the inner products of the  corresponding eigenvectors $v_i$. It the eigenvalues of $b_{ij}$ are not all positive, or not all negative, then the matrix receives a penalty equal to the number of positive or negative eigenvalues (whichever is lowest).\\

{\bf Item \ref{it:zero}\label{it:zero_ref}: The goal of absence of norm-minus-two vectors.} We want to give a penalty for every integer norm-minus-two vector (i.e.~a root) which is orthogonal to all positive-norm eigenvectors of $g \tilde g$ (the same discussion applies for $\tilde g g $, but here we only describe the procedure for $g \tilde g$).
Finding such vectors is a difficult task.
As discussed at the end of Section~\ref{sec:problem} we first need to establish if there are any such integer vectors at all (independent of their norm).
This can be achieved by factorizing the characteristic polynomial of $NdN^td$ over the integers,
\begin{equation}
p(\lambda) = \det(g \tilde g - \lambda \mathbb{1}) = \prod_r p_r(\lambda)^{m_r} \,,
\end{equation}
where each factor $p_r(\lambda)$ is irreducible over the integers.
Moreover, for each factor we define
\begin{equation}
N_r = \ker p_r(g \tilde g) \,.
\end{equation}
Here $p_r(g \tilde g)$ denotes the polynomial $p_r$ evaluated for the matrix $g \tilde g$.
Since the result is again an integer matrix, we are guaranteed to find an integer basis of $N_r$.
Eventually, we are only interested in those $N_r$ which contain only negative norm-square vectors, therefore we introduce
\begin{equation}
R_- = \left\{r : \text{$\left.(\cdot, \cdot)\right|_{N_r}$ is negative definite}\right\} \,.
\end{equation}
Clearly, we can write $N_-$ introduced in \eqref{eq:Nminus} as $N_- = \bigoplus_{r \in R_-} N_r$.
Therefore, as explained in Section~\ref{sec:problem} and Appendix~\ref{app:integervectors}, we are looking for integer vectors $v \in \mathbb{Z}^m$ which satisfy
\begin{equation}
p_-(g\tilde g) v \equiv \Biggl( \prod_{r \in R_-} p_r(g \tilde g ) \Biggr) v = 0 \,,\qquad \text{and}\qquad \left\|v\right\|^2 = -2 \,.
\end{equation}
To our understanding this quadratic Diophantine problem does not have a solution in general.
We can instead reformulate it into a lattice reduction problem. All integer vectors in the kernel of $\prod_{r \in R_-} p_r(g \tilde g)$
define a lattice (which we called $N_-$ earlier), so we need to find the smallest-norm vectors on this lattice.

The ordinary lattice reduction problem is an NP-hard problem, but there exists an algorithm that solves the problem, up to a potential ``defect'', in polynomial time.\footnote{We will gloss over the details of the defect, which is a central part of the LLL algorithm, since we will not make use of that feature.} This algorithm is called LLL after Lenstra, Lenstra, Lov\'asz \cite{Lenstra82factoringpolynomials}.
Given a set of integer vectors $\{v_k\}$, which span a lattice, the essential steps of the LLL algorithm are
\begin{enumerate}
  \item For the $k^{\text{th}}$ vector, $v_k$, in the kernel (arbitrarily ordered), starting at the second, replace $v_k$ with
  \begin{equation}
    v_k \leftarrow v_k + \lambda_{k,j} v_j
  \end{equation}
  iterating over all $j < k$, where $\lambda_{k,j} = -(v_k,v_j^\ast)/||v_j^\ast||^2$ rounded to an integer.\footnote{The asterisk refers to the corresponding vector after a Gram-Schmidt procedure (no normalization), and the Gram-Schmidt procedure is carried out at every redefinition of $v_k$.} \label{it:LLL1}
  \item If $v_k$ has a lower norm-square than $v_{k-1}$, switch their places and take a step back if possible ($k \leftarrow \max(k - 1, 2)$), else proceed to the next vector: $k \leftarrow k + 1$. \label{it:LLL2}
\end{enumerate}
Step \ref{it:LLL1} tries to lower the norm-squared of the $k^{\text{th}}$ vector using all earlier vectors. Step \ref{it:LLL2} orders the collection of vectors by increasing norm-squared and makes sure that the previous vectors are reprocessed if a new vector of shorter norm has been found.

We did not find a LLL-like algorithm that was implemented in a way that could be adapted to our problem however. This is because we use a custom inner product for our norm, and that norm is not positive definite. Most implementations do not support an input inner product, and we need to implement safe-guards to prevent singular expressions. We instead implement our own lattice-reduction algorithm that also takes into account several optimized features. The implementation can be found as part of the \texttt{bbsearch.jl} repository and is summarized in Appendix~\ref{app:LLL}.

In practice, however, we face yet another problem.
If we start with an arbitrary integer basis of $\ker p_-(g\tilde g)$, it is not guaranteed that every other vector in $N_- = \ker p_-(g\tilde g) \cap \mathbb{Z}^D$ can be expressed in terms of this basis with integer coefficients.
In general the coefficients can also be non-integer rationals.
However, such vectors will never be found by an LLL-type algorithm, as it only generates integer linear combinations of its input vectors.
In other words, it never leaves the lattice spanned by the input vectors.
Therefore, we must make sure that the vectors on which the LLL-type algorithm operates span not only a sublattice of $N_-$, but are a primitive basis of the full lattice $N_-$.
Otherwise, it might not be possible to find the shortest vector in $N_-$.
Schematically, the steps to find the shortest vectors in $N_-$ can be summarized as follows:
\begin{itemize}
\item Determine an integer basis $\{v_i\}$ of $\ker  p_-(g\tilde g) $.\footnote{We use \texttt{Nemo.jl}'s \cite{nemo} \texttt{nullspace\_right\_rational}-function that can return the nullspace as integer vectors.}
\item Compute a new basis $\{v'_i\}$ such that every integer vector in $\mathrm{span}_\mathbb{Q} \left(\{v_i\}\right)$ can be expressed with integer coefficients with respect to $\{v'_i\}$.
This basis will be a primitive basis of $N_-$.
An algorithm which performs this task is described in Appendix~\ref{app:rationalreduce}.
\item Use the algorithm from Appendix~\ref{app:LLL} to reduce the length of the basis vectors $\{v'_i\}$ as much as possible.
\end{itemize}
In our implementation we noticed that even if $N_-$ contains norm-minus-two vectors, sometimes none of them are found.
This is most prevalent if $\dim N_-$ is larger than a few.
To improve the performance of the search it can also help to perform the above steps on all or even just on some of the factors $p_r(g \tilde g)$ (with $r \in R_-$) of $p_-(g \tilde g)$ individually.
In practice, most of these factors will be polynomials of degree 0 or 1, which means that they directly correspond to integer eigenvalues; hence we can write them as $p_r(g \tilde g) = g \tilde g - e \mathbb{1}$ for $e$ an integer eigenvalue of $g \tilde g$.
Specifically, the three reduction paths we consider are:
\begin{enumerate}
  \item A basis of the kernel of $p_-(g\tilde g)$ is brought into Hermite normal form and normalized using the GCD. It is then processed by rational-reduce (described in Appendix \ref{app:rationalreduce}) brought again into Hermite normal form and finally lattice reduced.
  
    \item For each integer eigenvalue, $e$, the kernel of $(g \tilde g - e I)$ is GCD-normalized and put into Hermite normal form. The result is then, as above, put through rational-reduce, put into Hermite normal form and finally lattice reduced.
  
  \item For each integer eigenvalue, $e$, the kernel of $(g \tilde g - e I)$ is GCD-normalized, brought into Hermite normal form and then lattice reduced.
  
\end{enumerate}
In the order listed here, they are ranked according to the method that is most efficient at finding norm-minus-two vectors, and executed also in this order. Only the first reduction path is always taken, and the two others are only tested if the previous one did not find a norm-minus-two vector.

The matrix is then assigned a fitness value equal to the number of norm-minus-two vectors that were found. We also have implemented an option where the lattice reduction stops at the first norm-minus-two vector found. This can be useful if the weight for this penalty is set large (such that an order-one multiplicative factor does not make that much of a difference) to save time reducing the lattice. This comes at the cost of the DE not being able to solve this goal progressively as it does not have a way of distinguishing for example a matrix with four norm-minus-two vectors from one with three. We find that for small matrices this option is not really relevant, but that it helps for larger matrices.

We have not performed a worst-case scenario analysis of our implementation to determine its computational complexity, and since we have not implemented the feature of LLL that guarantees completion in polynomial time, we expect our algorithm not to have a polynomial scaling. This trade-off should mean longer execution time but less false-positive candidates. We have chosen to prioritize minimizing the number of false-positives here rather than trying to improve complexity.\\

{\bf Item \ref{it:tad}: The goal of minimal tadpole.} All of the previous goals and penalties must be zero for the matrix to be relevant. Since we are looking for the configuration with a minimal tadpole charge, we simply set the tadpole-charge penalty to
\begin{equation}
Q(N) = \tfrac12\mathrm{tr}(g \tilde g) \,.
\end{equation}

\subsection{Local search: ``the Spider''}\label{sec:spider}

Differential evolution is generally used to solve continuous problems, and the problem we are trying to solve is discrete, so the effectiveness of the algorithm will be limited. In order improve our chances of hitting the global minimum we also supplement the search with a local search.

In the late stages of the DE search, the diversity of the population will be very low, meaning that there are very few distinct integer matrices left over. Even at this stage, the DE will be able to continue explore the minima it has found by switching a few number of entries at a time between adjacent integers, albeit fairly slowly. At this stages we can design a more efficient algorithm that explores the vicinity of a certain matrix, where the concepts of ``vicinity'', ``local'' or ``closeness'' are related to how many entries differ between two matrices. Hence, two matrices are ``close'' if they differ by only a few entries and further apart if they differ by more entries.

To locally explore the minima around the final population generated by the DE, we use the following {\em Spider algorithm:}
\begin{itemize}
  \item {\bf filter:} Remove all matrices from the population that do not have all relevant penalties zero (as determined by the same fitness function we designed for the DE). Store these matrices in a filtered population.
  \item {\bf create:} For each matrix in the population being considered, generate two new matrices for each zero entry in the original matrix, replacing that zero with $\pm 1$. Once all matrices have been generated, filter the set to those matrices that has all relevant penalties zero. Add this set of matrices to the filtered population.
  \item {\bf kill:} For each matrix in the current set, generate one matrix for each non-zero entry where the selected entry is replaced with the next integer closer to zero. Filter again, and add to the filtered population.
\end{itemize}
One can find  versions of this Spider algorithm included as utility tools in the \texttt{bbsearch.jl} repository.

This Spider algorithm is then applied to matrices in batches. A batch is selected by choosing all matrices up to a certain tadpole charge such that the number of matrices is smaller or equal to the given batch-size. If there are more matrices with the lowest tadpole charge available than the batch-size, then the most recently generated matrices of that tadpole charge are selected. If they are still larger in number than the batch-size a batch-sized random sample of those matrices are selected for processing.

\begin{figure}[h]
  \begin{center}
    \includegraphics{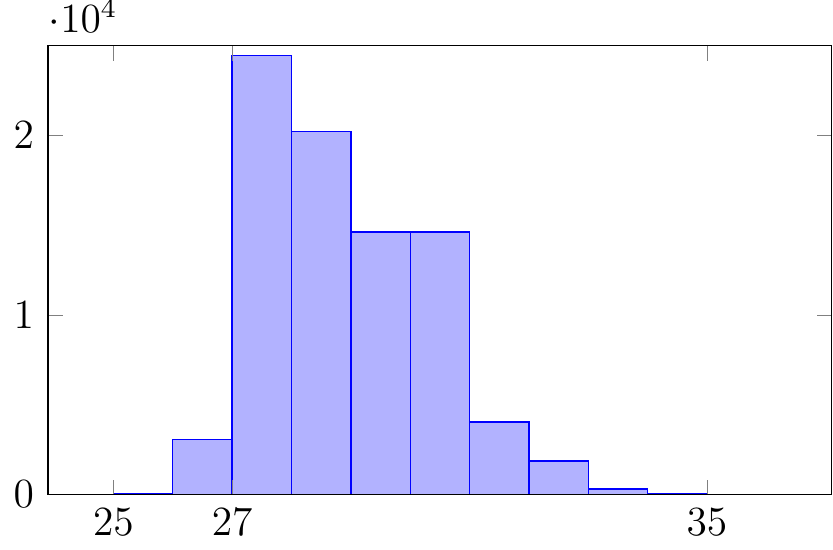}
  \end{center}
  \vspace{-1.5em}
  \caption{\it The distribution of tadpoles charges generated by a single step of the Spider algorithm when acting on a batch of 100 matrices out of a collection of $\mathit{45\,936}$ matrices with $\mathit{Q=25}$  (for the K3 lattice $2\, E_8 \oplus 3\, U$).  The lowest charge generated is $\mathit{Q=25}$ (33) and the highest $\mathit{Q=35}$ (2), and the most frequent charge is $\mathit{Q=27}$ $\mathit{(24\,4444)}$. In the next iteration of the Spider algorithm the 33 newly generated matrices with $\mathit{Q=25}$ will be processed, as the $\mathit{3\,079}$ matrices with $\mathit{Q=26}$ are too numerous for the batch-size (100). If no new $\mathit{Q=25}$ matrices are generated by them, a random sample of the remaining $\mathit{45\,836}$ matrices with $\mathit{Q=25}$ will be selected as the next batch for processing.
  }
  \label{fig:spidergen}
\end{figure}

By selecting the matrices for the batches this way, the Spider will generate and process matrices with lower tadpole charges, and prioritize processing the matrices generated by matrices of smaller tadpole charges. This is because what we found by experimenting that close to a matrix of a certain tadpole are matrices with similar and not too distant tadpoles. We illustrate an example of this phenomenon in Figure \ref{fig:spidergen}.

To further improve on the algorithm we also allow it to sample the complete set of matrices stored,\footnote{The implementation we have used does this every $10$ samplings.} because the minimal-tadpole matrices stored may represent a local minimum. To escape this local minimum we sometimes have to run the Spider algorithm not with the matrix with the lowest tadpole, but with a matrix of a higher tadpole.

While the process of filter-create-kill outlined above is deterministic, the Spider search as a whole is not. This because we need to limit the number of matrices we process by sampling, and to save memory and storage space, we need to truncate the number of stored matrices periodically.

\section{Results}\label{sec:toy}\label{sec:toy_models}
In this section we present our results and then discuss the performance of the search algorithm.

\subsection{Results of the DE plus Spider search}\label{sec:results}

In this section we present the results of our searches for various root lattices of dimensions $6 \leq D \leq 22$.
They are summarized in Table~\ref{tab:results}.
While all matrices are still too large to be explored exhaustively, some of them are small enough to be processed very quickly (both the differential evolution and the Spider converge on a minimum in a short amount of time).

\begin{table}[htb]
  \begin{center}
    \begin{tabular}{c|c|cccc|cc}
      lattice $\Lambda$    & $D$ & Runs & Population &  Distance & Time & $Q_\mathrm{min}$~(DE) & $Q_\mathrm{min}$~(Spider)\\
    \hline
    $3\,U$ & 6 & 8 & 500 & 2.5 & 12h & 5 (324) & 5 ($303\,290$) \\
    $A_4 \oplus U$ & 6 & 8 & 500 & 2.5 & 3h & 6 ($3\,256$) & 6 ($168\,348$) \\
    $D_4 \oplus U$ & 6 & 8 & 500 & 2.5 & 3h & 6 ($2\,938$) & 6 ($152\,377$)  \\
$A_4 \oplus 2\,U$  & 8 & 14 & 1\,000 & 2.5  & 12h & 7 (32) & 7 ($14\,745$)\\
$D_4 \oplus 2\, U$  & 8 & 14 & 1\,000 & 2.5  & 12h & 6 (299) & 6 ($57\,551$) \\
$E_6 \oplus U$ & 8 & 14 & 1\,000 & 2.5 & 12h & 9 ($4\,635$) & 9 ($224\,481$) \\
$A_4 \oplus 3\,U$  & 10 & 14 & 5\,000 & 2.5  & 48h & 9 (7) & 9 ($10\,204$) \\
$D_4 \oplus 3\,U$  & 10 & 14 & 5\,000 & 2.5  & 48h & 9 (282) & 9 ($69\,840$)  \\
    $E_8 \oplus U$ & 10  & 6    & $10\,000$        & $1.5$    & $6$h & 10 ($2$) & 10 ($47\,749$)\\
    $E_8 \oplus 2\,U$ & 12 & 6    & $5\,000$        & $1.5$    & $8$h & 14 ($1$) & 12 ($131\,426$)\\
    $E_8 \oplus 3\,U$ & 14 & 12   & $2\,500$        & $1.0$    & $10$h & 16 ($42$) & 13 ($1\,358$)\\
    $2\, E_6 \oplus 2\,U $  &16 & 15 & $5\,000$ & $1.5$ & 48h & 17 (31) & 14 (295)  \\
     $2\,E_8 \oplus U $ & 18 &  15    & $1\,000$        & $1.5$    & $36$h & 21 ($155$) & 20 ($42\,580$)\\
     $2\,E_8 \oplus 2\,U $ & 20 &  15    & $1\,000$        & $1.5$    & $72$h & 25 ($17$) & 21 ($66$) \\
     $2\,E_8 \oplus  3\,U$ & 22 & 16 & $1\,000$ & 0.7 & $72$h & 26 (5) & 25 ($100\,989$) \\
    \end{tabular}
  \end{center}
  \caption{\it Input data and results for various root lattices.
  ``Runs'' denotes the number of instances of the differential evolution search and ``Population'' the size of the population for each run.
  The ``Distance'' parameter is discussed in Subsection~\ref{sec:performance}.   ``Time'' is the runtime of each differential-evolution run.
  The last two columns list the minimal tadpole charge $Q_{\min}$ which we found with the differential evolution search (DE) and after processing the results with the Spider algorithm.
In the brackets we give the number of unique matrices with the respective minimal tadpole charge.}
  \label{tab:results}
\end{table}

In all searches we first started multiple instances of the differential evolution algorithm with varying parameters such as population size and runtime, depending on the dimension of the lattice.
Afterwards we used the Spider algorithm to explore the vicinity of the minima found by DE.
We observe that for lattices of smaller dimension, differential evolution and the Spider search converge to the same minimal tadpole charge, $Q$, while for larger lattices the Spider consistently manages to find matrices with smaller $Q$ than the DE. In all searches the Spider generated vastly more matrices with minimal $Q$ than the DE.

When increasing the lattice dimension, the performance of the DE becomes worse, since it fails to find a minimum within a reasonable amount of time. For small matrices, however, the situation is the opposite: The DE converges very quickly to a robust minimum which can be confirmed with the Spider search.
We take this as a clear indication that for each lattice $\Lambda$ there is a minimal charge, $Q_\mathrm{min}(\Lambda)$, which scales linearly with the dimension of the lattice.

\begin{figure}[hp]
  \begin{center}
    \begin{tabular}{cc}
      \includegraphics{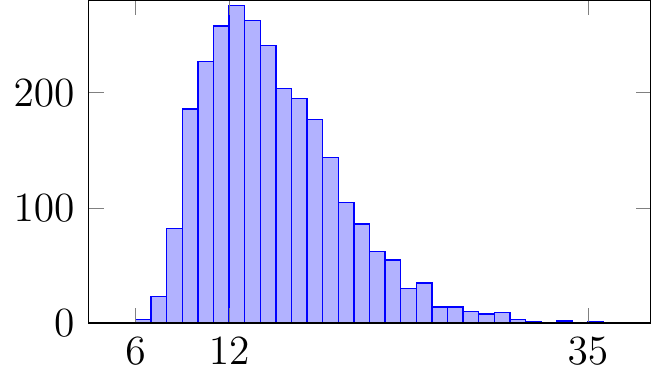} & \includegraphics{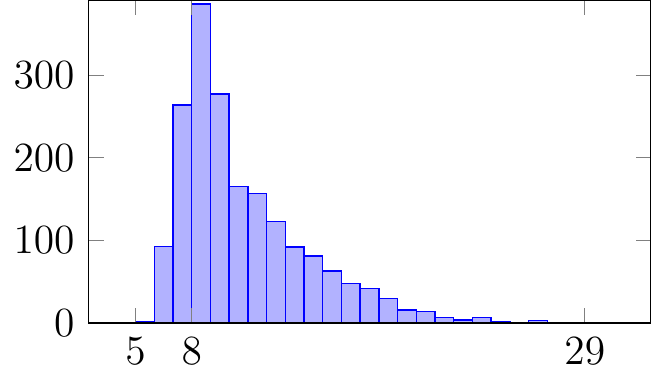} \\
      \includegraphics{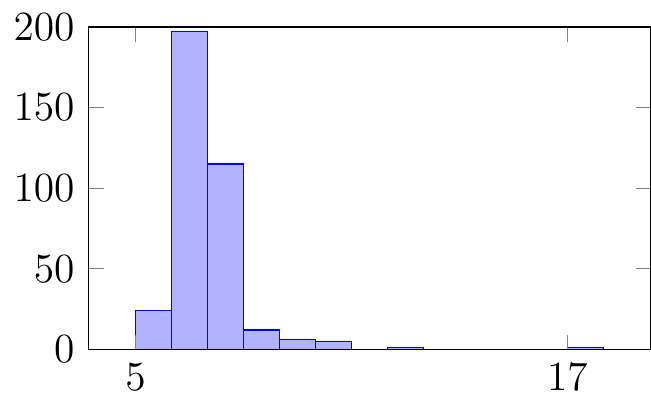} & \includegraphics{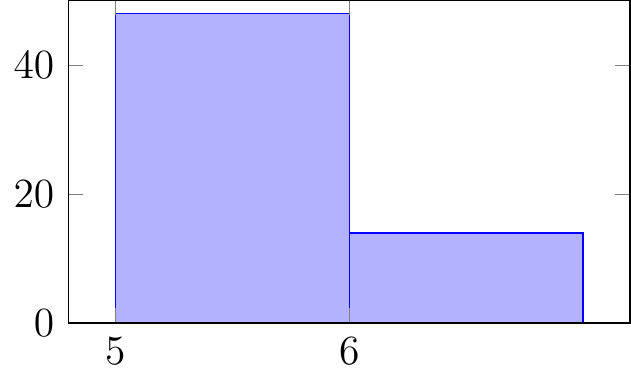}
    \end{tabular}
  \end{center}
  \vspace{-1.5em}
  \caption{\it Evolution of the population for $U \oplus U \oplus U$.
  Each histogram shows the distribution of tadpole charges $Q$ in the combined populations of 8 DE runs with a population of 500 each (counting unique integer matrices only).
 The data is taken after 60 seconds, 2 minutes, 5 minutes, and 30 minutes.}
  \label{fig:UUUruns}
\end{figure}

Let us illustrate this claim on the basis of the six-dimensional lattice $\Lambda = U \oplus U \oplus U$:
We started 8 instances of the differential evolution search with a population size of 500 each.
The evolution of these populations with time is depicted in Figure~\ref{fig:UUUruns}.
We show the distribution of the charge $Q$ in the cumulative population of all 8 runs after 1, 2, 5, and 30 minutes.
In the beginning the populations are very diverse and the charges seem to follow a smooth distribution, peaked around $Q = 12$.
Already after less than 2 minutes the first matrices with $Q = 5$ are found.
With increasing runtime the differential evolution algorithm homogenizes the populations and shifts the distribution of $Q$ towards smaller values, always bounded by the global minimum 5.
After around 50 minutes all matrices in the population have $Q = 5$ and even after 12 hours no single matrix with $Q < 5$ is found.
Post-processing the output of the differential evolution search with the Spider allowed us to generate a total of over 300 000 matrices with $Q = 5$ but again no matrices with smaller $Q$.
We therefore conclude that $Q_\mathrm{min} < 5$ is highly unlikely and we conjecture that
\begin{equation}
Q_\mathrm{min} \left(U \oplus U \oplus U\right) = 5 \,.
\end{equation}

\begin{figure}[hp]
  \begin{center}
    \begin{tabular}{cc}
      \includegraphics{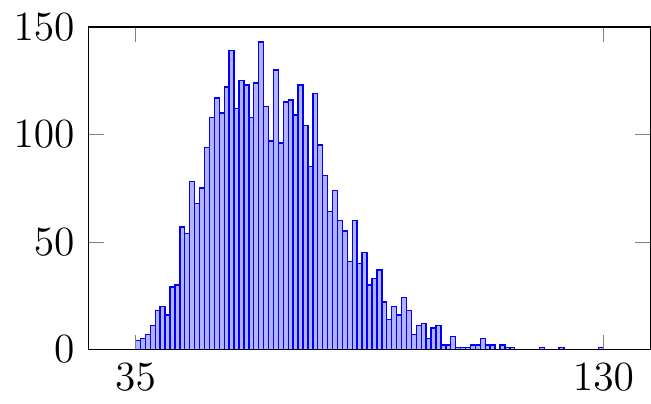} & \includegraphics{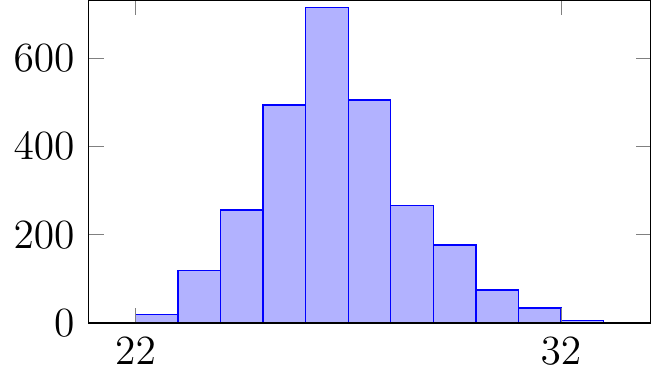} \\
      \includegraphics{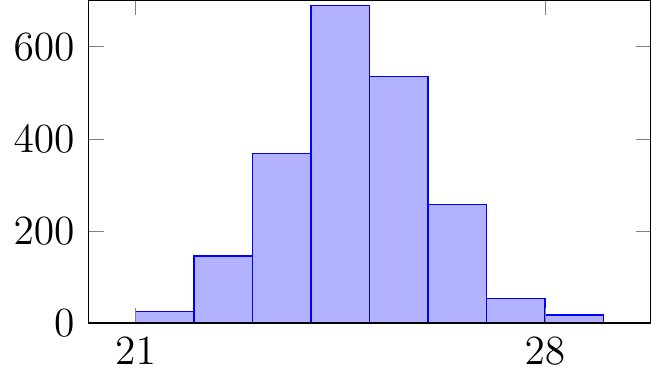} & \includegraphics{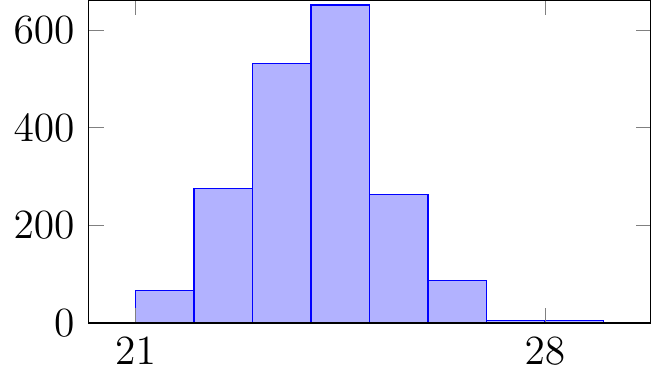}
    \end{tabular}
  \end{center}
  \vspace{-1.5em}
  \caption{\it Evolution of the population for $E_8 \oplus E_8 \oplus U$.
  Each histogram shows the distribution of tadpole charges $Q$ in the combined populations of 15 DE runs with a population of 1000 each.
 The data is taken after 20 minutes, 12 hours, 24 hours, and 36 hours.
  }
  \label{fig:E8E8Uruns}
\end{figure}

For larger dimensions it becomes increasingly more difficult to draw similar conclusions.
In Figure~\ref{fig:E8E8Uruns} we collected four snapshots of the populations for the 18-dimensional lattice $E_8 \oplus E_8 \oplus U$.
The data is from 14 instances of the DE with a population of 1000 each and was collected after 20 minutes, 12 hours, 24 hours and 36 hours.
Again, we observe that the diversity of the populations decreases with time and that the average value of $Q$ shifts to smaller values. However, this happens much more slowly than for the $U \oplus U \oplus U$ example.
In particular the minimal charge $Q = 21$ obtained with DE was only found after more than 20 hours and even after 36 hours most of the matrices in the population had still a larger charge.
Moreover, as shown by a subsequent Spider analysis, the DE did not manage to find the smallest possible charge.
With the Spider we found more than $40\,000$ matrices with $Q=20$.
The large amount of such matrices and the absence of matrices with smaller $Q$ gives a strong indication that here indeed $Q_\mathrm{min} = 20$.

\begin{figure}[hp]
  \begin{center}
    \begin{tabular}{ccc}

    \includegraphics{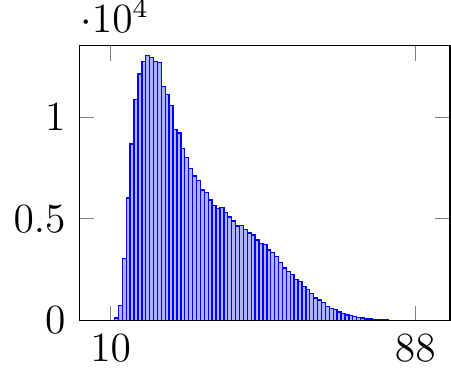} & \includegraphics{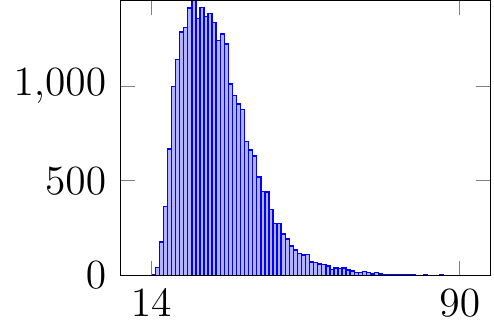} & \includegraphics{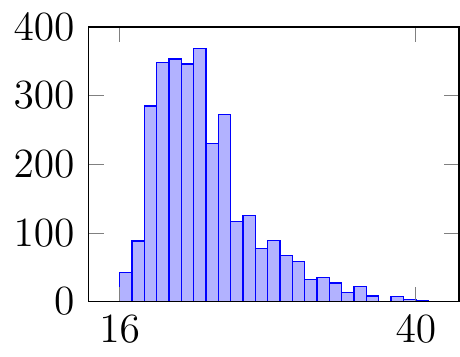} \\

    \includegraphics{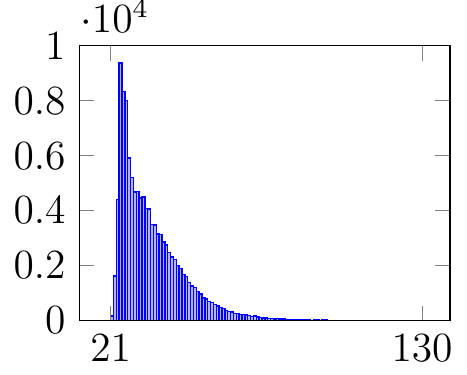} & \includegraphics{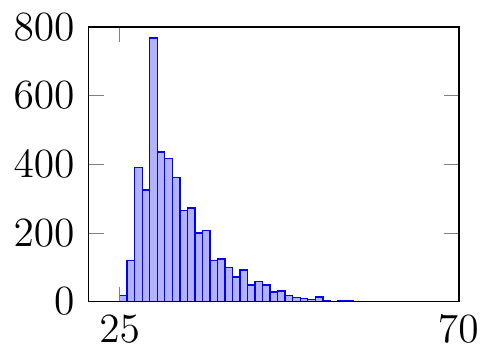} & \includegraphics{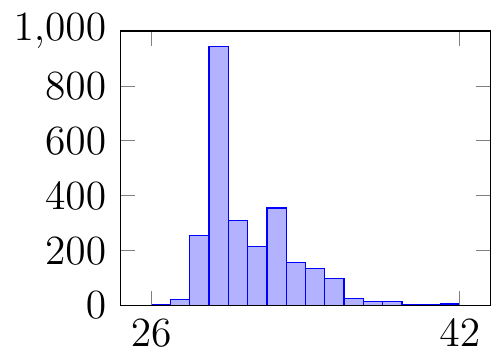}
    \end{tabular}
  \end{center}
  \caption{\it Tadpoles found by the DE for the lattices $E_8 \oplus U$, $E_8 \oplus 2\, U$, $E_8 \oplus 3\, U$, $2\, E_8 \oplus U$, $2\,E_8 \oplus 2\, U$, and $2\,E_8 \oplus 3\, U$ (for all all the flux configurations considered the relevant penalties are zero). More details can be found in Table \ref{tab:results}.}
  \label{fig:E8runs}
\end{figure}

In Figure \ref{fig:E8runs} we present the distribution of tadpoles from all snapshots taken during the runs with at least one $E_8$ factor. The distributions show a clear bias towards lower charges, as the DE attempts to find the minima. In the runs with two $E_8$ factors we see a much sharper fall-off towards lower charges, meaning that the DE has managed to generate a lot of matrices with low charges, indicating that we have not only let the search run for longer absolute time (see Table \ref{tab:results}) but also that it is longer in time relative to the increased complexity of the problem, going from a single $E_8$ to two.
It should be noted that these figures do not reveal much information about the true distribution of charges in the domain we consider, except that lower charges are indeed more rare.

For both large and small lattices, the differential evolution and the Spider algorithms show a clear universal linear scaling of $Q_\mathrm{min}$ with the dimension of the lattice
\begin{equation}
Q_\mathrm{min} \sim D \,.
\end{equation}
It is interesting to note that all our results are compatible with $Q_\mathrm{min} \geq D-2$.
Moreover, one can see that for small lattices $Q_\mathrm{min}(\Lambda)$ depends on the structure of the lattice, and not just on its dimension. It is an interesting question whether this dependence will persist in general for large lattices.

\subsection{Performance of the search algorithm}\label{sec:performance}

The naive way of approaching a problem similar to ours would be to either explore it exhaustively or randomly. The first option is very much unachievable even for small matrices, as even the smallest domain we consider ($U \oplus U \oplus U$, see Table \ref{tab:results}) contain $5^{6\times 6} \sim 10^{25}$ matrices. The second option, to explore this space randomly, performs much worse than differential evolution. The purpose of this subsection is to illustrate this phenomenon.

\begin{figure}
  \begin{center}
    \includegraphics[width=0.65\textwidth]{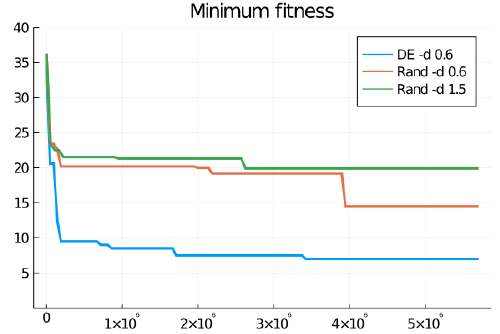}
  \end{center}
  \caption{\it A benchmark comparison between differential evolution and random search: fitness versus number of function evaluations. The blue line corresponding to how fitness have progressed for the differential evolution wins over those of the random search (even if we restrict the search space).}
  \label{fig:random_vs_de}
\end{figure}

We present three runs for the $E_8 \oplus U \oplus U$ matrix
\begin{enumerate}
  \item a DE run\footnote{As described in Section \ref{sec:algorithms}, we use \emph{adaptive} DE/rand/1/bin \emph{with radius limited sampling} for our DE searches, from \texttt{BlackBoxOptim.jl} \cite{Feldt2018}. The same implementation also includes a random sampling method that we use as a benchmark here.\label{fn:methods}} on the space $N^{ij} \in [-0.6,0.6]$ (this run has a \emph{distance} of $0.6$).
  \item a random sampling\footref{fn:methods} run on the space $N^{ij} \in [-1.5,1.5]$.
  \item a random sampling\footref{fn:methods} run on the space $N^{ij} \in [-0.6,0.6]$.
\end{enumerate}
keeping all other input to the fitness function and the algorithm fixed. The distance parameter determines in which domain each matrix entry, $N^{ij}$, is allowed to take values, at both initialization of the population (according to a random uniform distribution) and during the search. By these three runs we want to demonstrate what appears to be the generic behavior of algorithms applied to our problem:
\begin{itemize}
  \item The DE always outperforms random sampling
  \item The random sampling run over a more narrow search-space outperforms the one over the wider search-space
\end{itemize}
The first observation confirms at least to some degree that differential evolution is an approach suitable for our problem, as it is much more effective than just random sampling. In Figure \ref{fig:random_vs_de} we illustrate this by plotting the lowest-fitness member at regular intervals of the search, all three searches are rescaled as to have approximately the same number of total function evaluations at the final point.

The second observation tells us that we can improve a random search by including many zero entries in the matrix, which we have forced by choosing a smaller distance ($1.5$ versus $0.6$). This is consistent with what we find in the candidates with very low tadpoles. Hence, narrowing the search space benefits the random search, but still does not outperform the DE.

\section{Discussion}\label{sec:disc}\label{sec:discussion}

We divide this section into a discussion of the results we obtained, and a discussion of the algorithms we used.

\subsection{The results}

We have examined several lattices similar to that of K3$\times$K3 to support our claim that moduli stabilization at regular points of K3$\times$K3 compactifications of M-theory require a minimum tadpole charge  \cite{Bena:2020xrh}. Interestingly, this minimum charge, that we found to be 25, is larger than the maximum charge allowed by tadpole cancelation (24), and thus moduli stabilization always leads to  K3$\times$K3 manifolds with non-Abelian gauge groups.  The problem of moduli stabilization on K3$\times$K3 is particularly suited to an algorithmic treatment  as it can be translated into a set of algebraic conditions on a $22 \times 22$ matrix ($22=h^2(K3)$) that is quadratic in the flux quanta. Since $22$ is a considerably large number, certain runs take a very long time and one might wonder whether the minimum tadpole charge that we found is a mere artifact of a lack of running time.  This question was the main motivation for the present paper, were we have used the  algorithms to solve the same algebraic problem but with smaller matrices.

We applied our algorithms to several different lattices
which are characterized by their Gram matrix $d_{ij}$ representing an inner product of indefinite signature.%
\footnote{For actual compactifications, such as K3$\times$K3, this matrix corresponds to the intersection matrix of the middle cohomology.}
We analyzed examples in all even dimensions from 6 to 22.
All our examples are sums of roots lattices of simply laced Lie algebras ($A_n$, $D_n$ and $E_n$) and $U$.
This choice is of course motivated by the form of the K3 lattice $2\,E_8 \oplus 3\,U$.
Our main results are summarized in Table \ref{tab:results}.
We found that there is always a minimal tadpole charge, whose value is directly related to the dimension $D$ of the lattice.
The smallest charges found are always of order $D$.
In particular, all our examples are compatible with $Q_{\rm min}\ge D-2$, so one might wonder if this or a similar bound is universal.

Evidently, our algorithmic approach cannot give a mathematical proof that $Q_\mathrm{min}(\Lambda)$, for a given lattice, $\Lambda$, takes a certain value.
Nevertheless, as explained in Section~\ref{sec:results}, we found strong evidence that a non-trivial $Q_\mathrm{min}$ of the order of $\dim \Lambda$ always exists.
Translated in the language of flux compactifications, this means that one cannot stabilize all moduli at a generic point in moduli space with arbitarily small fluxes.
Establishing the actual value of $Q_\mathrm{min}(\Lambda)$ for a given lattice is of course a more difficult task, especially for lattices of large dimension.
It cannot be ruled out completely that the absence in our searches of any matrices with $Q$ smaller than a putative $Q_\mathrm{min}$ is merely a statistical effect.
If one takes a random sample of flux matrices, the spectrum of charges will follow a certain probability distribution favoring larger charges and strongly suppressing smaller charges.%
\footnote{This probability density could roughly be estimated in the following way: The number of flux configurations with a certain charge $Q$ is given by the number of lattice points in a shell of radius $Q$ and thickness 1. In the continuum limit, for $Q \gg 1$, the number of points is proportional to the volume of the shell.
Hence, for a spherical shell this would imply the na\"ive estimate $\rho(Q) \sim Q^{D^2}$.
This, however, does not take into account that we are dealing with a space with non-definite inner product.
Moreover, our constraints on moduli stabilization and absence of singularities further reduce the number of admissible matrices. Finally, this simple counting argument breaks down for small values of $Q$, when the effects of flux quantization become important. We thank D.~Junghans for interesting discussions on this argument.}
This effect will increase with the dimension of $\Lambda$, which explains why finding $Q_\mathrm{min}$ becomes more difficult for larger lattices.
Therefore, establishing $Q_\mathrm{min}$ from a purely random search appears to require a sample size which is large enough to contain at least a few instances with $Q = Q_\mathrm{min}$ with sufficiently large statistical probability.

However, as illustrated in Section~\ref{sec:performance}, our algorithms are vastly more efficient than a random search.
The charge distribution of the matrices found by the differential evolution and Spider algorithms is shifted strongly towards smaller $Q$, hence allowing for smaller sample sizes.
It is outside the scope of this work to quantify the effect of the search algorithms on the charge distribution.
Nevertheless, we believe that we have generated enough matrices to make our results for $Q_\mathrm{min}$ rather trustworthy. In particular, for the 22-dimensional K3-lattice we generated a very large number of matrices with $Q = 25$ relatively easy and fast, but failed to generate even a single flux matrix with $Q < 25$.

It would be desirable to further explore the properties of the statistical distribution of fluxes of the K3$\times$K3 compactification and to better quantify the performance of the differential evolution and spider algorithms.
Clearly, in this context a comparison with the general results on the distribution of flux vacua of \cite{Douglas:2003um,Denef:2004ze} would be very interesting.
There also appear to be promising connections to the theory of random matrices.
Finally, it does not seem to be completely unrealistic to obtain $Q_\mathrm{min}$ by purely analytic arguments, at least for some special lattices.
In particular there exist connections between our problem and number theory,\footnote{We thank Cumrun Vafa for suggesting this.} see for example \cite{Moore:1998zu, Moore:1998pn, Gukov:2002nw, Wendland:2003ma, Moore:2004fg, rizov2005complex, Chen:2005gm, Schimmrigk:2006dy, Rohde:2010tv, ito2018supersingular, Benjamin:2018mdo, valloni2018complex, Kachru:2020sio, Yang:2020gwr, Banerjee:2020szx, Kanno:2020kxr}, as also suggested by the exposition in Appendix~\ref{app:integervectors}.

\subsection{The algorithms}

To solve our tadpole-minimization problem we used two different search methods: Differential Evolution (DE) and the Spider. The DE approach has worked quite well, even though it is not directly intended for use on discrete problems. The DE pinpoints fairly easily a good region where minima are located,  but can be slow in finding the actual minima, especially for larger lattices. Thus, the DE converges fairly consistently on a region of low tadpoles, only a few units away from where the minima are, but has trouble producing the actual minima without extra help.

This help is provided by our Spider algorithm, which explores the area in which the DE settles, by turning matrix entries on and off. The Spider has proven very good at progressively improve the tadpole minima generated by the DE. Combining the DE and the Spider and performing several searches, we find what appear to be the global minima.

The success of DE in pinpointing the area where the minima are is a feature that we believe to be of general interest. Indeed, more modern AI techniques, primarily those that fall under the topic of machine learning (ML), often require lot of computational time for training before deciding if the methods chosen are suitable for the problem. The relative advantage of DE, or other meta-heuristic or stochastic optimization algorithms, is that they can directly be applied to a problem when a fitness function has been built. It would be interesting to explore the limits of the application of these algorithms to the problems we face.

In particular, it would be interesting to compare our algorithms to the ML approach used for example in \cite{Halverson:2019tkf} to solve a system of coupled non-linear Diophantine equations. In ML the resulting neural network is trained to be an algorithm that solves a  particular problem, and it would be interesting to see how this algorithm performs compared to other meta-heuristic or deterministic algorithms, and not just random walks.

Since ML techniques have been used to solve games, it would be interesting to try to explore the problem considered here in some similar way. For example, our Spider algorithm is very similar to a game of Solitaire, where a move is to switch on and off individual entries and improve the tadpole charge, while still satisfying all targets. The goal of this modified Solitaire game would be to find the minimal tadpole, using these moves. One difference is that instead of facing an opponent, a machine would have to learn how to play the perfect game.

It would also be interesting to see if there are other ways in which one can formulate our minimization problem. In our current formulation, it does not appear that our question is even in the class of NP-hard problems, as we cannot verify a candidate matrix with certainty in polynomial time, because of the NP-hard lattice reduction problem. Exploring this further may yield some more suitable formulation of the problem and even better ways of finding solutions.

\vspace{0.6cm}
\noindent {\bf Acknowledgements:} We would like to thank Andreas Braun, Thomas Grimm, Jim Halverson, Arthur Hebecker, Lena Heijkenskj{\"o}ld, Daniel Junghans, Daniel Mayerson, Ruben Monten, Jakob Moritz and Cumrun Vafa for interesting discussions. The work of I.B., M.G.~and S.L.~was supported in part by the ANR grant Black-dS-String ANR-16-CE31-0004-01, by the ERC Grants 772408 ``Stringlandscape'' and 787320 ``QBH Structure'', by the John Templeton Foundation grant 61149 and by the NSF Grant PHY-1915071, PHY-1748958 and PHY-1607611.  The work of J.B.~was supported by the MIUR-PRIN contract 2015MP2CX4002 ``Non-perturbative aspects of gauge theories and strings''.

We are grateful to all contributors of the open-source projects that made this work possible, which includes: Jupyter notebooks \cite{Kluyver2016jupyter}, Julia \cite{Julia-2017} and packages \texttt{BlackboxOptim.jl} \cite{Feldt2018}, \texttt{Nemo.jl} \cite{nemo}, \texttt{Polynomials.jl}, \texttt{GenericLinearAlgebra.jl}, \texttt{Plots.jl}, \texttt{Measures.jl}, \texttt{StatsBase.jl},\!  \texttt{UnicodePlots.jl},\! \texttt{BenchmarkTools.jl},  \texttt{ArgParse.jl}, \texttt{ProgressMeter.jl}, and all of their dependencies.


\appendix


\section{Integer linear combinations of eigenvectors}\label{app:integervectors}

In this Appendix we discuss under which conditions a subset of the eigenvectors of an integer matrix has an integer vector in its span.

In Section~\ref{sec:K3K3} we explained that the K3$\times$K3 compactifications develop an orbifold singularity if one of the integer matrices $g \tilde g$ and $\tilde g g$ with $g$ and $\tilde g$ defined in \eqref{eq:gmatrix} has the following property:
There is a root of the K3 lattice \eqref{intK3} which is orthogonal to all eigenvectors with positive norm square.
A necessary condition for this is that there is a linear combination of the eigenvectors with negative norm square which is integer (see the discussion in Section~\ref{sec:toy_models} for more details).
In this Appendix we will outline the underlying mathematics of this question.

Let $M \in \mathbb{Z}^{D \times D}$ be a diagonalizable $D \times D$ matrix with integer entries and $\lambda_i$ and $v_i$ ($i = 1, \dots, D$) its eigenvalues and the corresponding eigenvectors.
In general neither the eigenvalues nor the eigenvectors are integer.
We want to know under which circumstances there exists a subset $\{v_k\, |\, k = 1, \dots, p < d\}$ of the eigenvectors such that there exists a linear combination that has only integer entries:
\begin{equation}\label{eq:intlincombination}
\sum_{i=1}^p a^i v_i \in \mathbb{Z^d} \,,\qquad a^i \in \mathbb{C} \,.
\end{equation}

The two limits of this questions are easy to understand.
For $p = d$ equation  \eqref{eq:intlincombination} is always true since $\mathbb{Z}^D$ is contained in the (complex) span of all eigenvectors, which is $\mathbb{C}^D$.
In the opposite limit, $p = 1$, equation \eqref{eq:intlincombination} is also true  if there exists an eigenvalue $\lambda \in \mathbb{Z}$; its corresponding eigenvector, $v$, is given by the solution of
\begin{equation}
\left(M - \lambda \mathbb{1}\right) v = 0 \,.
\end{equation}
This is a linear equation with integer coefficients and admits therefore an integer solution.
We would now like to understand whether there exist solutions to this problem away from these limits.

The eigenvalues $\lambda_i$ are the $D$ (complex) roots of the characteristic polynomial
\begin{equation}
p(\lambda) = \det \left(M - \lambda \mathbb{1}\right) = \lambda^D + c_{d-1} \lambda^{D-1} + \dots + c_0 \,,
\end{equation}
where the coefficients, $c_i$, are integers.
This polynomial is not always irreducible over the integers. When it is not, we can decompose it into a product of smaller polynomials with integer coefficients:
\begin{equation}\label{eq:polyndecomp}
p(\lambda) = \prod_r p_r(\lambda) \,,
\end{equation}
where each of the factors is irreducible over the integers.
This means that the polynomials
\begin{equation}
p_r(\lambda) = \lambda^{d_r} + c^{(r)}_{d_r - 1} \lambda^{d_r -1} + \dots +  c^{(r)}_{0} \,,\qquad c^{(r)}_i \in \mathbb{Z}
\end{equation}
cannot be further decomposed into a product of non-constant integer polynomials of smaller rank.
Here we denote the degree of $p_r(\lambda)$ by $d_r < D$.

Let us now focus on one of these polynomials, $p_r(\lambda)$.
In principle, this polynomial can occur multiple times in the factorization \eqref{eq:polyndecomp}, so we denote its multiplicity by $m_r$. Furthermore,
$p_r(\lambda)$ has $d_r$ complex roots $\lambda^{(r)}_i$.
By construction these roots are eigenvalues of $M$, so we denote the corresponding eigenvectors by $v^{(r)}_i$.
Notice, that if $m_r > 1$ there are more eigenvectors than eigenvalues, since for each eigenvalue, $\lambda^{(r)}_i$, there are $m_r$ independent eigenvectors.
Next, we consider the matrix
\begin{equation}
p_r(M) = M^{d_r} + c^{(r)}_{d_r - 1} M^{d_r -1} + \dots +  c^{(r)}_{0} \mathbb{1} \,.
\end{equation}
For every eigenvector, $v^{(r)}_i$, it satisfies
\begin{equation}
p_r(M) v^{(r)}_i = p_r(\lambda^{(r)}_i) \, v^{(r)}_i = 0 \,.
\end{equation}
Therefore, $\dim \ker p_r(M) \geq m_r d_r$.
This inequality can only hold for every $r$ if it is saturated and we have
\begin{equation}
\ker p_r(M) = \mathrm{span} \left\{v_i^{(r)}\,\bigl|\, i = 1, \dots, m_r d_r\right\} \,.
\end{equation}
On the other hand, $p_r(M)$ is an integer matrix and thus its kernel admits an integer basis $\{b^{(r)}_i \,,\, i =  1, \dots, m_r d_r\} \in \mathbb{Z}^{d \times m_r d_r}$.
Therefore, there exist $m_r p_r$ independent linear combinations
\begin{equation}
\sum_{i=1}^{m_r d_r} a^i v^{(r)}_i \in \mathbb{Z^d} \,.
\end{equation}
Notice that, if $m_r > 1$, we can always redefine the eigenvectors in such a way that the sum contains only $p_r$ non-vanishing terms.
To conclude, for each irreducible factor $p_r(\lambda)$ there exist $p_r$ independent eigenvectors such that one can form $d_r$ independent integer vectors from their linear combinations.
We can finally express the matrix $M$ with respect to the integer basis $\bigcup_r \{b^{(r)}_i\}$ to block-diagonalize it over the integers.
The characteristic polynomials of the individual blocks will be the irreducible polynomials $p_r(\lambda)$.

Let us also look at the reverse situation and assume there is a subset of eigenvectors $\{\tilde v_i\}$, $i = 1, \dots, \tilde d < D$, such that there is a linear combination
\begin{equation}
b_1 = \sum_{i = 1}^{\tilde d} a^i \tilde  v_i \in \mathbb{Z}^d \,.
\end{equation}
However, denoting the corresponding eigenvalues by $\tilde \lambda_i$, we can show iteratively that all
\begin{equation}
b_n \equiv \sum_{i = 1}^{\tilde d} a^i \tilde\lambda_i^n  \tilde v_i = M b_{n-1} \in \mathbb{Z}^d \,.
\end{equation}
On the other hand, at most $\tilde d$ of these vectors can be linearly independent. Therefore, there exists a linear relation between the $b_i$, $i = 0, \dots, \tilde d$.
This shows that the $\tilde\lambda_i$ are roots of an integer polynomial of degree $\tilde d$ and therefore $p(\lambda)$ must be reducible.
Alternatively, one can extend the $b_i$ ($i = 0, \dots, \tilde d$) to an integer basis of $\mathbb{C}^D$ and express $M$ with respect to this basis.
This brings $M$ into block diagonal form which shows that $p(\lambda)$ is reducible.

This shows that there exists an integer linear combination of $p < D$ eigenvectors if and only if the characteristic polynomial of $M$ is reducible and if in its factorization there is an irreducible factor of degree p.

Clearly, the $p=1$ example mentioned above, where there is a solution if one of the eigenvalues is integer, can be seen as limit of our general result, obtained when the corresponding irreducible factor $p_r(\lambda)$ has degree $d_r = 1$.

The roots of any of the irreducible factors $p_r$ form an algebraic extension of the rationals and we are thus dealing with linear algebra over these algebraic fields.
The discussion here therefore seems to indicate that there are potentially very interesting connections of our problem to algebraic number theory, which should be elaborated on elsewhere.


\section{Lattice Reduction}\label{app:latticereduction}

In this Appendix we present some details of the algorithms which we use to determine if there are integer norm-minus-two vectors (roots) in the kernel of an integer matrix.
In \ref{app:rationalreduce} we outline a simple algorithm which computes a primitive basis of the kernel of an integer matrix.
In \ref{app:LLL} we discuss our adaptation of the LLL algorithm which tries to find the shortest vector in a given lattice.

\subsection{Rational reduction}\label{app:rationalreduce}

Let $M \in \mathbb{Z}^{D \times D}$ be an integer matrix and $v_i \in \mathbb{Z}^D$ ($i = 1,  \dots, k$; $k = \dim \ker M$) a basis of the kernel of $M$.
The vectors $\{v_i\}$ are called a primitive basis (of the lattice $\ker M \cap \mathbb{Z}^D$) if every integer vector in the kernel of $M$ can be expressed as a linear combination o the $v_i$ with integer coefficients:
\begin{equation}
\mathrm{span}_\mathbb{Z} (v_1, \dots, v_k) =  \ker M \cap \mathbb{Z}^D \equiv \ker_\mathbb{Z} M \,.
\end{equation}
Not every integer basis is necessarily primitive as it can be possible to obtain integer vectors as rational linear combinations of integer vectors.

Here, we give a simple algorithm which generates for a set of integer vectors, $\{v_i\}$, a new, primitive basis of $\mathrm{span}_\mathbb{Q} (\{v_i\}) \cap \mathbb{Z}^D$.
We start by writing the $v_i$ as the columns of a matrix $V$:
\begin{equation}
V \equiv \begin{pmatrix}
\vert & & \vert\\
v_1 &\hspace{-0.1cm} \cdots \hspace{-0.1cm}& v_k \\
\vert & & \vert \\
\end{pmatrix} \,.
\end{equation}
The algorithm then works as follows:
\begin{enumerate}
\item Bring $V$ into column Hermite normal form and normalize all columns by dividing them by the GCD of their entries. If $V$ was already in this form skip to the next step. The matrix $V$ is now in column echelon form.
\item Iterate over $i = 1, \dots, k$: \\[0.5ex]
Introduce the reduced matrix
\begin{equation}
V_i \equiv \begin{pmatrix}
\vert & & \vert\\
v_i &\hspace{-0.1cm} \cdots \hspace{-0.1cm}& v_k \\
\vert & & \vert \\
\end{pmatrix} \,.
\end{equation}
Determine all non-zero elements of $v_i$ such that all other entries in the same row of $V_i$ are zero ($v_{i,j} \neq 0$ and $v_{m,j} = 0$ for $m > i$).
Denote the GCD of these elements by $Q$ and decompose $Q$ into prime factors. For each prime factor $p$ of $Q$ do:
\begin{enumerate}
\item Compute a basis $n_j \in\left(\mathbb{Z}_p\right)^D$ of the kernel of $V_i$
in $\mathbb{Z}_p \equiv \mathbb{Z} / p \mathbb{Z}$. (This means each $n_j$ satisfies $V_i \, n_j = 0 \mod p$.)
\item Bring the matrix
\begin{equation}
\begin{pmatrix}
\vert & & \vert\\
n_1 &\hspace{-0.1cm} \cdots \hspace{-0.1cm}& n_l \\
\vert & & \vert \\
\end{pmatrix}
\end{equation}
into column Hermite normal form.
Divide each $n_j$ by its leading element (its first non-zero entry).
These operations are done mod $p$.
\item For each $n_j$, denote the position of its leading element by $l_i$ (i.e.~$n_{j, l_j} = 1$ and $n_{j, m} = 0$ for $m > l_j$). Replace $v_{i + l_j -1}$ by $V_i \, n_j / p$.
\end{enumerate}
\end{enumerate}

\subsection{LLL}\label{app:LLL}

Given a set of integer vectors, $\{ v_k \}$, which span a lattice, the ``shortest vector problem'' asks what for the length of the shortest non-zero vector in the this lattice.
The LLL algorithm \cite{Lenstra82factoringpolynomials} (outlined in \ref{ssec:fitness}) gives an approximate solution to this problem in polynomial time.
Here, we use the following adaptation of LLL:

\begin{enumerate}
  \item Iterating over all vectors $v_k$ for $k \in [1,n]$, replace $v_k$ with a new vector
  \begin{equation}
    v_k \leftarrow v_k + \lambda^i v_i \,.
  \end{equation}
  where $\lambda^i$ is the nearest integer to the solution of
  \begin{equation}\label{eq:mmin}
    d_{ij} \lambda^j + c_i = 0;\quad d_{ij} = (v_i,v_j)\,,\ c_i = (v_k,v_i)\,,
  \end{equation}
  and $i,j \in [1,n]$ except $i,j = k$ and any $l$ with vector $v_l$ that makes $d_{ij}$ singular.
  Notice that the left hand side of \eqref{eq:mmin} is the derivative of
    \begin{equation}\label{eq:newvector}
   \|v_k + \lambda^i v_i \|^2 =  \|v_k\|^2 + 2\lambda^i c_i + \lambda^i \lambda^j d_{ij}\,,
  \end{equation}
  with respect to $\lambda^i$.
  Hence a solution of \eqref{eq:mmin} extremizes the length of the new vector.
  Existence and uniqueness of $\lambda^i$ is guaranteed if $d_{ij}$ is non-degenerate.
  If it is moreover positive (negative) definite, the extremum of \eqref{eq:newvector} will be indeed a minimum (maximum).

  \item For all vectors $v_i$ with $i \neq k$ that have $||v_i|| = 0$ but $(v_i,v_k) \neq 0$ we replace $v_k$ with
  \begin{equation}
    v_k \leftarrow v_k + v_i \kappa_i;\quad \kappa_i = \frac{2-||v_k||^2}{2(v_k,v_i)}\,,
  \end{equation}
  for the $v_i$ that leaves the new $v_k$ closest to $||v_k||^2 = 2$.
  \item Repeat the above procedure until no vector in the collection has changed.
\end{enumerate}

Both the $\lambda_i$ and $\kappa_i$ must be integer (or an appropriate rational) to make the result integer. While we have chosen rounding for most problems we have studied,\footnote{If we do not use rounding, our algorithm calculates \emph{all} combinations of rounding to largest/smallest integer, to get the best possible solution. For large matrices with many vectors in the kernel this however becomes too slow.} that is not always the one that gives the result closest to $||v_k||^2 = 2$, but it performs better than using any of the \texttt{floor}/\texttt{ceil} functions.

If we compare our Item 1.~above to the conventional LLL, we find that the two (disregarding rounding) solve the same analytical problem. Since we opted to solve Eq.~(\ref{eq:mmin}) in the most intuitive way -- by matrix inversions -- LLL would probably have a computational complexity advantage to our implementation, even if the LLL performed more iterations finish. The two algorithms also differ in that LLL rounds at every iteration.



\newpage

\bibliographystyle{JHEP}
\bibliography{refs-tadpole}

\providecommand{\href}[2]{#2}\begingroup\raggedright\begin{thebibliography}{10}

\bibitem{Bena:2020xrh}
I.~Bena, J.~Bl\r{a}b\"ack, M.~Gra\~na, and S.~L\"ust, {\it {The Tadpole
  Problem}},  \href{http://arxiv.org/abs/2010.10519}{{\tt arXiv:2010.10519}}.

\bibitem{Braun:2020jrx}
A.~P. Braun and R.~Valandro, {\it {$G_4$ Flux, Algebraic Cycles and Complex
  Structure Moduli Stabilization}},
  \href{http://arxiv.org/abs/2009.11873}{{\tt arXiv:2009.11873}}.

\bibitem{Giryavets:2003vd}
A.~Giryavets, S.~Kachru, P.~K. Tripathy, and S.~P. Trivedi, {\it {Flux
  compactifications on Calabi-Yau threefolds}},  {\em JHEP} {\bf 04} (2004)
  003, [\href{http://arxiv.org/abs/hep-th/0312104}{{\tt hep-th/0312104}}].

\bibitem{Demirtas:2019sip}
M.~Demirtas, M.~Kim, L.~Mcallister, and J.~Moritz, {\it {Vacua with Small Flux
  Superpotential}},  {\em Phys. Rev. Lett.} {\bf 124} (2020), no.~21 211603,
  [\href{http://arxiv.org/abs/1912.10047}{{\tt arXiv:1912.10047}}].

\bibitem{Collinucci:2008pf}
A.~Collinucci, F.~Denef, and M.~Esole, {\it {D-brane Deconstructions in IIB
  Orientifolds}},  {\em JHEP} {\bf 02} (2009) 005,
  [\href{http://arxiv.org/abs/0805.1573}{{\tt arXiv:0805.1573}}].

\bibitem{Dasgupta:1999ss}
K.~Dasgupta, G.~Rajesh, and S.~Sethi, {\it {M theory, orientifolds and G -
  flux}},  {\em JHEP} {\bf 08} (1999) 023,
  [\href{http://arxiv.org/abs/hep-th/9908088}{{\tt hep-th/9908088}}].

\bibitem{Aspinwall:2005ad}
P.~S. Aspinwall and R.~Kallosh, {\it {Fixing all moduli for M-theory on
  K3xK3}},  {\em JHEP} {\bf 10} (2005) 001,
  [\href{http://arxiv.org/abs/hep-th/0506014}{{\tt hep-th/0506014}}].

\bibitem{Braun:2008pz}
A.~P. Braun, A.~Hebecker, C.~Ludeling, and R.~Valandro, {\it {Fixing D7 Brane
  Positions by F-Theory Fluxes}},  {\em Nucl. Phys.} {\bf B815} (2009)
  256--287, [\href{http://arxiv.org/abs/0811.2416}{{\tt arXiv:0811.2416}}].

\bibitem{Storn:1996}
R.~Storn, {\it {On the usage of differential evolution for function
  optimization}},  in {\em Biennial Conference of the North American Fuzzy
  Information Processing Society}, NAFIPS, pp.~519--523, 1996.

\bibitem{Storn:1997}
R.~Storn and K.~Price, {\it {Differential Evolution – A Simple and Efficient
  Heuristic for global Optimization over Continuous Spaces}},  {\em Journal of
  Global Optimization} {\bf 11} (1997), no.~4 341--359.

\bibitem{Price:2005}
K.~Price, R.~M. Storn, and J.~A. Lampinen, {\em {Differential Evolution: A
  Practical Approach to Global Optimization}}.
\newblock Springer-Verlag Berlin Heidelberg, Springer, 2005.

\bibitem{Engelbrecht}
A.~P. Engelbrecht, {\em {Computational Intelligence: An Introduction}}.
\newblock John Wiely \& Sons, Ltd, Wiely, 2007.

\bibitem{Blaback:2013ht}
J.~Bl\r{a}b\"ack, U.~Danielsson, and G.~Dibitetto, {\it {Fully stable dS vacua
  from generalised fluxes}},  {\em JHEP} {\bf 08} (2013) 054,
  [\href{http://arxiv.org/abs/1301.7073}{{\tt arXiv:1301.7073}}].

\bibitem{Damian:2013dq}
C.~Damian, L.~R. Diaz-Barron, O.~Loaiza-Brito, and M.~Sabido, {\it {Slow-Roll
  Inflation in Non-geometric Flux Compactification}},  {\em JHEP} {\bf 06}
  (2013) 109, [\href{http://arxiv.org/abs/1302.0529}{{\tt arXiv:1302.0529}}].

\bibitem{Damian:2013dwa}
C.~Damian and O.~Loaiza-Brito, {\it {More stable de Sitter vacua from S-dual
  nongeometric fluxes}},  {\em Phys. Rev. D} {\bf 88} (2013), no.~4 046008,
  [\href{http://arxiv.org/abs/1304.0792}{{\tt arXiv:1304.0792}}].

\bibitem{Blaback:2013fca}
J.~Bl\r{a}b\"ack, U.~Danielsson, and G.~Dibitetto, {\it {Accelerated Universes
  from type IIA Compactifications}},  {\em JCAP} {\bf 03} (2014) 003,
  [\href{http://arxiv.org/abs/1310.8300}{{\tt arXiv:1310.8300}}].

\bibitem{Blaback:2013qza}
J.~Bl\r{a}b\"ack, D.~Roest, and I.~Zavala, {\it {De Sitter Vacua from
  Nonperturbative Flux Compactifications}},  {\em Phys. Rev. D} {\bf 90}
  (2014), no.~2 024065, [\href{http://arxiv.org/abs/1312.5328}{{\tt
  arXiv:1312.5328}}].

\bibitem{Abel:2014xta}
S.~Abel and J.~Rizos, {\it {Genetic Algorithms and the Search for Viable String
  Vacua}},  {\em JHEP} {\bf 08} (2014) 010,
  [\href{http://arxiv.org/abs/1404.7359}{{\tt arXiv:1404.7359}}].

\bibitem{Ruehle:2017mzq}
F.~Ruehle, {\it {Evolving neural networks with genetic algorithms to study the
  String Landscape}},  {\em JHEP} {\bf 08} (2017) 038,
  [\href{http://arxiv.org/abs/1706.07024}{{\tt arXiv:1706.07024}}].

\bibitem{Cole:2019enn}
A.~Cole, A.~Schachner, and G.~Shiu, {\it {Searching the Landscape of Flux Vacua
  with Genetic Algorithms}},  {\em JHEP} {\bf 11} (2019) 045,
  [\href{http://arxiv.org/abs/1907.10072}{{\tt arXiv:1907.10072}}].

\bibitem{AbdusSalam:2020ywo}
S.~AbdusSalam, M.~Cicoli, F.~Quevedo, P.~Shukla, and S.~Abel, {\it {A
  systematic approach to K\"ahler moduli stabilisation}},  {\em JHEP} {\bf 08}
  (2020), no.~08 047, [\href{http://arxiv.org/abs/2005.11329}{{\tt
  arXiv:2005.11329}}].

\bibitem{CaboBizet:2020cse}
N.~Cabo~Bizet, C.~Damian, O.~Loaiza-Brito, D.~K.~M. Pe\~na, and J.~Monta\~nez
  Barrera, {\it {Testing Swampland Conjectures with Machine Learning}},  {\em
  Eur. Phys. J. C} {\bf 80} (2020), no.~8 766,
  [\href{http://arxiv.org/abs/2006.07290}{{\tt arXiv:2006.07290}}].

\bibitem{Betzler:2019kon}
P.~Betzler and E.~Plauschinn, {\it {Type IIB flux vacua and tadpole
  cancellation}},  {\em Fortsch. Phys.} {\bf 67} (2019), no.~11 1900065,
  [\href{http://arxiv.org/abs/1905.08823}{{\tt arXiv:1905.08823}}].

\bibitem{He:2017aed}
Y.-H. He, {\it {Deep-Learning the Landscape}},
  \href{http://arxiv.org/abs/1706.02714}{{\tt arXiv:1706.02714}}.

\bibitem{Carifio:2017bov}
J.~Carifio, J.~Halverson, D.~Krioukov, and B.~D. Nelson, {\it {Machine Learning
  in the String Landscape}},  {\em JHEP} {\bf 09} (2017) 157,
  [\href{http://arxiv.org/abs/1707.00655}{{\tt arXiv:1707.00655}}].

\bibitem{Carifio:2017nyb}
J.~Carifio, W.~J. Cunningham, J.~Halverson, D.~Krioukov, C.~Long, and B.~D.
  Nelson, {\it {Vacuum Selection from Cosmology on Networks of String
  Geometries}},  {\em Phys. Rev. Lett.} {\bf 121} (2018), no.~10 101602,
  [\href{http://arxiv.org/abs/1711.06685}{{\tt arXiv:1711.06685}}].

\bibitem{Hashimoto:2018ftp}
K.~Hashimoto, S.~Sugishita, A.~Tanaka, and A.~Tomiya, {\it {Deep learning and
  the AdS/CFT correspondence}},  {\em Phys. Rev. D} {\bf 98} (2018), no.~4
  046019, [\href{http://arxiv.org/abs/1802.08313}{{\tt arXiv:1802.08313}}].

\bibitem{Wang:2018rkk}
Y.-N. Wang and Z.~Zhang, {\it {Learning non-Higgsable gauge groups in 4D
  F-theory}},  {\em JHEP} {\bf 08} (2018) 009,
  [\href{http://arxiv.org/abs/1804.07296}{{\tt arXiv:1804.07296}}].

\bibitem{Bull:2018uow}
K.~Bull, Y.-H. He, V.~Jejjala, and C.~Mishra, {\it {Machine Learning CICY
  Threefolds}},  {\em Phys. Lett. B} {\bf 785} (2018) 65--72,
  [\href{http://arxiv.org/abs/1806.03121}{{\tt arXiv:1806.03121}}].

\bibitem{Demirtas:2018akl}
M.~Demirtas, C.~Long, L.~McAllister, and M.~Stillman, {\it {The Kreuzer-Skarke
  Axiverse}},  {\em JHEP} {\bf 04} (2020) 138,
  [\href{http://arxiv.org/abs/1808.01282}{{\tt arXiv:1808.01282}}].

\bibitem{Constantin:2018hvl}
A.~Constantin and A.~Lukas, {\it {Formulae for Line Bundle Cohomology on
  Calabi-Yau Threefolds}},  {\em Fortsch. Phys.} {\bf 67} (2019), no.~12
  1900084, [\href{http://arxiv.org/abs/1808.09992}{{\tt arXiv:1808.09992}}].

\bibitem{Klaewer:2018sfl}
D.~Klaewer and L.~Schlechter, {\it {Machine Learning Line Bundle Cohomologies
  of Hypersurfaces in Toric Varieties}},  {\em Phys. Lett. B} {\bf 789} (2019)
  438--443, [\href{http://arxiv.org/abs/1809.02547}{{\tt arXiv:1809.02547}}].

\bibitem{Halverson:2018cio}
J.~Halverson and F.~Ruehle, {\it {Computational Complexity of Vacua and
  Near-Vacua in Field and String Theory}},  {\em Phys. Rev. D} {\bf 99} (2019),
  no.~4 046015, [\href{http://arxiv.org/abs/1809.08279}{{\tt
  arXiv:1809.08279}}].

\bibitem{Mutter:2018sra}
A.~M\"utter, E.~Parr, and P.~K.~S. Vaudrevange, {\it {Deep learning in the
  heterotic orbifold landscape}},  {\em Nucl. Phys. B} {\bf 940} (2019)
  113--129, [\href{http://arxiv.org/abs/1811.05993}{{\tt arXiv:1811.05993}}].

\bibitem{Altman:2018zlc}
R.~Altman, J.~Carifio, J.~Halverson, and B.~D. Nelson, {\it {Estimating
  Calabi-Yau Hypersurface and Triangulation Counts with Equation Learners}},
  {\em JHEP} {\bf 03} (2019) 186, [\href{http://arxiv.org/abs/1811.06490}{{\tt
  arXiv:1811.06490}}].

\bibitem{He:2018jtw}
Y.-H. He, {\it {The Calabi-Yau Landscape: from Geometry, to Physics, to
  Machine-Learning}},  \href{http://arxiv.org/abs/1812.02893}{{\tt
  arXiv:1812.02893}}.

\bibitem{Cole:2018emh}
A.~Cole and G.~Shiu, {\it {Topological Data Analysis for the String
  Landscape}},  {\em JHEP} {\bf 03} (2019) 054,
  [\href{http://arxiv.org/abs/1812.06960}{{\tt arXiv:1812.06960}}].

\bibitem{Bull:2019cij}
K.~Bull, Y.-H. He, V.~Jejjala, and C.~Mishra, {\it {Getting CICY High}},  {\em
  Phys. Lett. B} {\bf 795} (2019) 700--706,
  [\href{http://arxiv.org/abs/1903.03113}{{\tt arXiv:1903.03113}}].

\bibitem{Halverson:2019kna}
J.~Halverson, C.~Long, B.~Nelson, and G.~Salinas, {\it {Axion reheating in the
  string landscape}},  {\em Phys. Rev. D} {\bf 99} (2019), no.~8 086014,
  [\href{http://arxiv.org/abs/1903.04495}{{\tt arXiv:1903.04495}}].

\bibitem{Hashimoto:2019bih}
K.~Hashimoto, {\it {AdS/CFT correspondence as a deep Boltzmann machine}},  {\em
  Phys. Rev. D} {\bf 99} (2019), no.~10 106017,
  [\href{http://arxiv.org/abs/1903.04951}{{\tt arXiv:1903.04951}}].

\bibitem{Halverson:2019tkf}
J.~Halverson, B.~Nelson, and F.~Ruehle, {\it {Branes with Brains: Exploring
  String Vacua with Deep Reinforcement Learning}},  {\em JHEP} {\bf 06} (2019)
  003, [\href{http://arxiv.org/abs/1903.11616}{{\tt arXiv:1903.11616}}].

\bibitem{He:2019vsj}
Y.-H. He and S.-J. Lee, {\it {Distinguishing elliptic fibrations with AI}},
  {\em Phys. Lett. B} {\bf 798} (2019) 134889,
  [\href{http://arxiv.org/abs/1904.08530}{{\tt arXiv:1904.08530}}].

\bibitem{Brodie:2019dfx}
C.~R. Brodie, A.~Constantin, R.~Deen, and A.~Lukas, {\it {Machine Learning Line
  Bundle Cohomology}},  {\em Fortsch. Phys.} {\bf 68} (2020), no.~1 1900087,
  [\href{http://arxiv.org/abs/1906.08730}{{\tt arXiv:1906.08730}}].

\bibitem{Ashmore:2019wzb}
A.~Ashmore, Y.-H. He, and B.~A. Ovrut, {\it {Machine Learning
  Calabi\textendash{}Yau Metrics}},  {\em Fortsch. Phys.} {\bf 68} (2020),
  no.~9 2000068, [\href{http://arxiv.org/abs/1910.08605}{{\tt
  arXiv:1910.08605}}].

\bibitem{Parr:2019bta}
E.~Parr and P.~K.~S. Vaudrevange, {\it {Contrast data mining for the MSSM from
  strings}},  {\em Nucl. Phys. B} {\bf 952} (2020) 114922,
  [\href{http://arxiv.org/abs/1910.13473}{{\tt arXiv:1910.13473}}].

\bibitem{Halverson:2019vmd}
J.~Halverson, M.~Plesser, F.~Ruehle, and J.~Tian, {\it {K\"ahler Moduli
  Stabilization and the Propagation of Decidability}},  {\em Phys. Rev. D} {\bf
  101} (2020), no.~4 046010, [\href{http://arxiv.org/abs/1911.07835}{{\tt
  arXiv:1911.07835}}].

\bibitem{Halverson:2020opj}
J.~Halverson and C.~Long, {\it {Statistical Predictions in String Theory and
  Deep Generative Models}},  {\em Fortsch. Phys.} {\bf 68} (2020), no.~5
  2000005, [\href{http://arxiv.org/abs/2001.00555}{{\tt arXiv:2001.00555}}].

\bibitem{Ruehle:2020jrk}
F.~Ruehle, {\it {Data science applications to string theory}},  {\em Phys.
  Rept.} {\bf 839} (2020) 1--117.

\bibitem{Parr:2020oar}
E.~Parr, P.~K.~S. Vaudrevange, and M.~Wimmer, {\it {Predicting the orbifold
  origin of the MSSM}},  {\em Fortsch. Phys.} {\bf 68} (2020), no.~5 2000032,
  [\href{http://arxiv.org/abs/2003.01732}{{\tt arXiv:2003.01732}}].

\bibitem{Otsuka:2020nsk}
H.~Otsuka and K.~Takemoto, {\it {Deep learning and k-means clustering in
  heterotic string vacua with line bundles}},  {\em JHEP} {\bf 05} (2020) 047,
  [\href{http://arxiv.org/abs/2003.11880}{{\tt arXiv:2003.11880}}].

\bibitem{Deen:2020dlf}
R.~Deen, Y.-H. He, S.-J. Lee, and A.~Lukas, {\it {Machine Learning String
  Standard Models}},  \href{http://arxiv.org/abs/2003.13339}{{\tt
  arXiv:2003.13339}}.

\bibitem{Krippendorf:2020gny}
S.~Krippendorf and M.~Syvaeri, {\it {Detecting Symmetries with Neural
  Networks}},  \href{http://arxiv.org/abs/2003.13679}{{\tt arXiv:2003.13679}}.

\bibitem{He:2020eva}
Y.-H. He, E.~Hirst, and T.~Peterken, {\it {Machine-Learning Dessins d'Enfants:
  Explorations via Modular and Seiberg-Witten Curves}},
  \href{http://arxiv.org/abs/2004.05218}{{\tt arXiv:2004.05218}}.

\bibitem{Akutagawa:2020yeo}
T.~Akutagawa, K.~Hashimoto, and T.~Sumimoto, {\it {Deep Learning and AdS/QCD}},
   {\em Phys. Rev. D} {\bf 102} (2020), no.~2 026020,
  [\href{http://arxiv.org/abs/2005.02636}{{\tt arXiv:2005.02636}}].

\bibitem{Bao:2020nbi}
J.~Bao, S.~Franco, Y.-H. He, E.~Hirst, G.~Musiker, and Y.~Xiao, {\it {Quiver
  Mutations, Seiberg Duality and Machine Learning}},  {\em Phys. Rev. D} {\bf
  102} (2020), no.~8 086013, [\href{http://arxiv.org/abs/2006.10783}{{\tt
  arXiv:2006.10783}}].

\bibitem{He:2020bfv}
Y.-H. He, {\it {Calabi-Yau Spaces in the String Landscape}},
  \href{http://arxiv.org/abs/2006.16623}{{\tt arXiv:2006.16623}}.

\bibitem{Bies:2020gvf}
M.~Bies, M.~Cveti\v{c}, R.~Donagi, L.~Lin, M.~Liu, and F.~Ruehle, {\it {Machine
  Learning and Algebraic Approaches towards Complete Matter Spectra in 4d
  F-theory}},  \href{http://arxiv.org/abs/2007.00009}{{\tt arXiv:2007.00009}}.

\bibitem{Erbin:2020srm}
H.~Erbin and R.~Finotello, {\it {Inception Neural Network for Complete
  Intersection Calabi-Yau 3-folds}},
  \href{http://arxiv.org/abs/2007.13379}{{\tt arXiv:2007.13379}}.

\bibitem{Erbin:2020tks}
H.~Erbin and R.~Finotello, {\it {Machine learning for complete intersection
  Calabi-Yau manifolds: a methodological study}},
  \href{http://arxiv.org/abs/2007.15706}{{\tt arXiv:2007.15706}}.

\bibitem{Demirtas:2020dbm}
M.~Demirtas, L.~McAllister, and A.~Rios-Tascon, {\it {Bounding the
  Kreuzer-Skarke Landscape}},  \href{http://arxiv.org/abs/2008.01730}{{\tt
  arXiv:2008.01730}}.

\bibitem{He:2020lbz}
Y.-H. He and A.~Lukas, {\it {Machine Learning Calabi-Yau Four-folds}},
  \href{http://arxiv.org/abs/2009.02544}{{\tt arXiv:2009.02544}}.

\bibitem{Parr:2020dbl}
E.~Parr, {\em {Machine Learning in String Theory}}.
\newblock PhD thesis, Munich, Tech. U., 2020.

\bibitem{Cole:2020ktw}
A.~E. Cole, {\em {Identifying and Exploiting Structure in Cosmological and
  String Theoretic Data}}.
\newblock PhD thesis, U. Wisconsin, Madison (main), 2020.

\bibitem{Jejjala:2020wcc}
V.~Jejjala, D.~K. Mayorga~Pena, and C.~Mishra, {\it {Neural Network
  Approximations for Calabi-Yau Metrics}},
  \href{http://arxiv.org/abs/2012.15821}{{\tt arXiv:2012.15821}}.

\bibitem{proceedingsAjtai:1997}
M.~Ajtai, {\it {The Shortest Vector Problem is NP-hard for Randomized
  Reductions}},  in {\em Electronic Colloquium on Computational Complexity},
  1997.

\bibitem{proceedingsAjtai:1998}
M.~Ajtai, {\it {The Shortest Vector Problem is NP-hard for Randomized
  Reductions}},  in {\em Proceedings 30th Annual ACM Symposium on Theory of
  Computing}, 1998.

\bibitem{Lenstra82factoringpolynomials}
A.~K. Lenstra, H.~W. Lenstra, and L.~Lovasz, {\it Factoring polynomials with
  rational coefficients},  {\em MATH. ANN} {\bf 261} (1982) 515--534.

\bibitem{bbsearch}
J.~Bl{\aa}b{\"a}ck, ``Julia framework to simplify analysis of problems via the
  use of blackboxoptim.jl.'' \url{https://gitlab.com/johanbluecreek/bbsearch}.

\bibitem{Braun:2010ff}
A.~P. Braun, {\em {F-Theory and the Landscape of Intersecting D7-Branes}}.
\newblock PhD thesis, Heidelberg U., 2010.
\newblock \href{http://arxiv.org/abs/1003.4867}{{\tt arXiv:1003.4867}}.

\bibitem{Feldt2018}
R.~Feldt, ``Blackboxoptim.jl.''
  \url{https://github.com/robertfeldt/BlackBoxOptim.jl}, 2018.

\bibitem{Julia-2017}
J.~Bezanson, A.~Edelman, S.~Karpinski, and V.~B. Shah, {\it Julia: A fresh
  approach to numerical computing},  {\em SIAM {R}eview} {\bf 59} (2017), no.~1
  65--98.

\bibitem{nemo}
C.~Fieker, W.~Hart, T.~Hofmann, and F.~Johansson, {\it Nemo/hecke: Computer
  algebra and number theory packages for the julia programming language},  in
  {\em Proceedings of the 2017 ACM on International Symposium on Symbolic and
  Algebraic Computation}, ISSAC '17, (New York, NY, USA), pp.~157--164, ACM,
  2017.

\bibitem{Douglas:2003um}
M.~R. Douglas, {\it {The Statistics of string / M theory vacua}},  {\em JHEP}
  {\bf 05} (2003) 046, [\href{http://arxiv.org/abs/hep-th/0303194}{{\tt
  hep-th/0303194}}].

\bibitem{Denef:2004ze}
F.~Denef and M.~R. Douglas, {\it {Distributions of flux vacua}},  {\em JHEP}
  {\bf 05} (2004) 072, [\href{http://arxiv.org/abs/hep-th/0404116}{{\tt
  hep-th/0404116}}].

\bibitem{Moore:1998zu}
G.~W. Moore, {\it {Attractors and arithmetic}},
  \href{http://arxiv.org/abs/hep-th/9807056}{{\tt hep-th/9807056}}.

\bibitem{Moore:1998pn}
G.~W. Moore, {\it {Arithmetic and attractors}},
  \href{http://arxiv.org/abs/hep-th/9807087}{{\tt hep-th/9807087}}.

\bibitem{Gukov:2002nw}
S.~Gukov and C.~Vafa, {\it {Rational conformal field theories and complex
  multiplication}},  {\em Commun. Math. Phys.} {\bf 246} (2004) 181--210,
  [\href{http://arxiv.org/abs/hep-th/0203213}{{\tt hep-th/0203213}}].

\bibitem{Wendland:2003ma}
K.~Wendland, {\it {On Superconformal field theories associated to very
  attractive quartics}},  in {\em {Les Houches School of Physics: Frontiers in
  Number Theory, Physics and Geometry}}, pp.~223--244, 2007.
\newblock \href{http://arxiv.org/abs/hep-th/0307066}{{\tt hep-th/0307066}}.

\bibitem{Moore:2004fg}
G.~W. Moore, {\it {Strings and Arithmetic}},  in {\em {Les Houches School of
  Physics: Frontiers in Number Theory, Physics and Geometry}}, pp.~303--359,
  2007.
\newblock \href{http://arxiv.org/abs/hep-th/0401049}{{\tt hep-th/0401049}}.

\bibitem{rizov2005complex}
J.~Rizov, {\it Complex multiplication for k3 surfaces},
  \href{http://arxiv.org/abs/math/0508018}{{\tt math/0508018}}.

\bibitem{Chen:2005gm}
M.~Chen, {\it {Complex multiplication, rationality and mirror symmetry for
  Abelian varieties}},  {\em J. Geom. Phys.} {\bf 58} (2008) 633--653,
  [\href{http://arxiv.org/abs/math/0512470}{{\tt math/0512470}}].

\bibitem{Schimmrigk:2006dy}
R.~Schimmrigk, {\it {The Langlands program and string modular K3 surfaces}},
  {\em Nucl. Phys. B} {\bf 771} (2007) 143--166,
  [\href{http://arxiv.org/abs/hep-th/0603234}{{\tt hep-th/0603234}}].

\bibitem{Rohde:2010tv}
J.~C. Rohde, {\it {Some Mirror partners with Complex multiplication}},  {\em
  Commun. Num. Theor. Phys.} {\bf 4} (2010) 597--607,
  [\href{http://arxiv.org/abs/1006.0807}{{\tt arXiv:1006.0807}}].

\bibitem{ito2018supersingular}
K.~Ito, {\it On the supersingular reduction of k3 surfaces with complex
  multiplication},  2018.

\bibitem{Benjamin:2018mdo}
N.~Benjamin, S.~Kachru, K.~Ono, and L.~Rolen, {\it {Black holes and class
  groups}},  \href{http://arxiv.org/abs/1807.00797}{{\tt arXiv:1807.00797}}.

\bibitem{valloni2018complex}
D.~Valloni, {\it Complex multiplication and brauer groups of k3 surfaces},
  2018.

\bibitem{Kachru:2020sio}
S.~Kachru, R.~Nally, and W.~Yang, {\it {Supersymmetric Flux Compactifications
  and Calabi-Yau Modularity}},  \href{http://arxiv.org/abs/2001.06022}{{\tt
  arXiv:2001.06022}}.

\bibitem{Yang:2020gwr}
W.~Yang, {\it {K3 mirror symmetry, Legendre family and Deligne's conjecture for
  the Fermat quartic}},  {\em Nucl. Phys. B} {\bf 963} (2021) 115303,
  [\href{http://arxiv.org/abs/2004.00820}{{\tt arXiv:2004.00820}}].

\bibitem{Banerjee:2020szx}
A.~Banerjee and G.~W. Moore, {\it {Hyperk\"ahler isometries of K3 surfaces}},
  {\em JHEP} {\bf 12} (2020) 193, [\href{http://arxiv.org/abs/2009.11769}{{\tt
  arXiv:2009.11769}}].

\bibitem{Kanno:2020kxr}
K.~Kanno and T.~Watari, {\it {W=0 Complex Structure Moduli Stabilization on
  CM-type K3 x K3 Orbifolds:---Arithmetic, Geometry and Particle Physics---}},
  \href{http://arxiv.org/abs/2012.01111}{{\tt arXiv:2012.01111}}.

\bibitem{Kluyver2016jupyter}
T.~Kluyver, B.~Ragan-Kelley, F.~P{\'e}rez, B.~Granger, M.~Bussonnier,
  J.~Frederic, K.~Kelley, J.~Hamrick, J.~Grout, S.~Corlay, P.~Ivanov, D.~Avila,
  S.~Abdalla, and C.~Willing, {\it Jupyter notebooks -- a publishing format for
  reproducible computational workflows},  in {\em Positioning and Power in
  Academic Publishing: Players, Agents and Agendas} (F.~Loizides and
  B.~Schmidt, eds.), pp.~87 -- 90, IOS Press, 2016.

\end{thebibliography}\endgroup

\end{document}